\documentclass[journal]{IEEEtran}
\ifCLASSINFOpdf
\else
\fi

\usepackage{cite}
\usepackage{amsfonts}
\usepackage{amsmath}
\usepackage[pdftex]{graphicx}
\usepackage{amsmath,amssymb,amsfonts}
\usepackage{algorithmic}
\usepackage{graphicx}
\usepackage{booktabs} 
\usepackage{textcomp}
\usepackage{xcolor}
\usepackage{float}

\usepackage{hyperref}
\hypersetup{
            colorlinks=true,
            linkcolor=blue,
            anchorcolor=blue,
            citecolor=blue,
            urlcolor=blue
            }

\usepackage{amsmath}
\newtheorem{theorem}{Theorem}

\begin{document}
%
\title{Joint Computation Offloading and Resource Allocation for Maritime MEC with Energy Harvesting}
%
%
%

\author{Zhen~Wang,~\IEEEmembership{Student Member,~IEEE,}
        Bin~Lin,~\IEEEmembership{Senior Member,~IEEE,}
        Qiang (John) Ye,~\IEEEmembership{Senior Member,~IEEE,}
        Yuguang~Fang,~\IEEEmembership{Fellow,~IEEE,}
        and~Xiaoling~Han,~\IEEEmembership{Student Member,~IEEE}
\thanks{Copyright (c) 20xx IEEE. Personal use of this material is permitted. However, permission to use this material for any other purposes must be obtained from the IEEE by sending a request to pubs-permissions@ieee.org.}
\thanks{\emph{(Corresponding author: Bin Lin.) }}
\thanks{Zhen Wang is with the Department of Information Science and Technology of Dalian Maritime University, Dalian, China, and Communication Engineering of Dalian Neusoft University of Information, Dalian, China. E-mail: wangzhen\underline{~}jsj@neusoft.edu.cn.}
\thanks{Bin Lin and Xiaoling Han are with the Department of Information Science and Technology of Dalian Maritime University, Dalian, China. E-mail: binlin@dlmu.edu.cn, xiaolinghan@dlmu.edu.cn.}
\thanks{Qiang (John) Ye is with the Department of Electrical and Software Engineering, Schulich School of Engineering, University of Calgary, 2500 University Drive NW, Calgary, AB T2N 1N4. Email: qiang.ye@ucalgary.ca.}
\thanks{Yuguang Fang is with the Department of Computer Science City University of Hong Kong
Tat Chee Avenue, Kowloon, Hong Kong. E-mail: my.fang@cityu.edu.hk.}
\thanks{This manuscript has been accepted by IEEE Internet of Things Journal, DOI: 10.1109/JIOT.2024.3371049. }
}

\maketitle

\begin{abstract}
In this paper, we establish a multi-access edge computing (MEC)-enabled sea lane monitoring network (MSLMN) architecture with energy harvesting (EH) to support dynamic ship tracking, accident forensics, and anti-fouling through real-time maritime traffic scene monitoring. Under this architecture, the computation offloading and resource allocation are jointly optimized to maximize the long-term average throughput of MSLMN. Due to the dynamic environment and unavailable future network information, we employ the Lyapunov optimization technique to tackle the optimization problem with large state and action spaces and formulate a stochastic optimization program subject to queue stability and energy consumption constraints. We transform the formulated problem into a deterministic one and decouple the temporal and spatial variables to obtain asymptotically optimal solutions. Under the premise of queue stability, we develop a joint computation offloading and resource allocation (JCORA) algorithm to maximize the long-term average throughput by optimizing task offloading, subchannel allocation, computing resource allocation, and task migration decisions. Simulation results demonstrate the effectiveness of the proposed scheme over existing approaches.
\end{abstract}


%
\IEEEpeerreviewmaketitle

\section{Introduction}
%
%
%
%
 \IEEEPARstart{N}{owadays}, more than 80$\%$ of the total international cargo transportation is carried over the sea. The increasing maritime activities such as maritime transportation, sea lane
monitoring, and marine resource extraction lead to high demand for maritime information exchange \cite{ref1,ref2,ref3,ref4,ref5}. On one hand, the density of marine vessels (especially the ones near big harbors) is increasing tremendously, which puts forward higher requirements for inter-vessel connections through advanced wireless communication technologies (e.g., LTE) for information dissemination. On the other hand, the information sharing from diversified applications to realize the ``smart ocean" often requires high data rates and low transmission latency. 
For example, in the sea lane monitoring scenario, it is necessary to quickly analyze and process the collected images/videos of sea lanes, to make real-time prediction on vessel behaviors and provide accurate navigation assistance and efficient navigation services for vessels.
Generally, to achieve more intelligent and responsive event monitoring, video perception, and dynamic tracking, a large amount of data needs to be sensed/collected and processed promptly which inevitably increases the on-vessel computation burden. 
However, due to geographical restrictions, the allocation of communication and computing resources from terrestrial networks to support maritime services is often limited, which poses significant challenges when supporting massive communication and computing demands. It is imperative to develop a more efficient networking and computing solution to achieve better performance for massive maritime data transmission and task processing.

Multi-access edge computing (MEC) technology remains an effective solution, with which the computation-intensive tasks can be offloaded, through wireless communication networks, to onshore/offshore more powerful edge servers for processing with high efficiency \cite{ref6,ref7,ref8,ref9}. In traditional cellular systems such as 5G and beyond (5G+), a multi-tier networking architecture, consisting of a macro cell base station (MBS)  underlaid by multiple small cell base stations (SBS),  has been proposed to cope with the massive data and enhance the communication coverage\cite{ref10,ref11,ref12}. Both SBSs and MBSs are deployed with MEC servers and the data traffic can be scheduled towards the MBS and SBS flexibly.
Due to such attractive features, the applications of MEC to maritime network scenarios have also 
attracted great attention from academia. An MEC-based space-air-ground integrated maritime communication network is proposed to support various maritime applications \cite{ref13}. For computation-intensive applications at sea, a voyage-based computation offloading scheme is proposed with edge nodes deployed on vessels to provide MEC services for nearby users \cite{ref14}. A computation offloading scheme based on an improved Hungarian algorithm for multi-vessels \cite{ref15} and a reinforcement learning based intelligent task offloading algorithm \cite{ref16} have been proposed to enhance the maritime MEC performance, such as on delay and energy consumption. Many existing proposals  consider a single layer of MEC platform and optimize the computation offloading efficiency between local and edge processing various performance indicators (such as latency and energy consumption). 
However, the impact of communication and computing resource scheduling on computation offloading performance still needs further research, especially in maritime MEC with dynamic environment and limited resources, and how to efficiently leverage MEC to support the sea lane monitoring requirements still demands further investigation.
 
Meanwhile, considering limited accessibility to the ground power grid for maritime communication networks, how to resolve energy supply for communications and computing for long term operations is highly challenging and energy harvesting (EH) still remains a promising solution\cite{ref17}, \cite{ref18}. Most existing studies mainly focus on the energy management for EH-enabled end devices. The use of EH to power small cell base stations (SBS) has also been proposed to jointly optimize caching and user-base station association issues \cite{ref19}. Considering the characteristics of a marine communication environment, most maritime communication infrastructures are away from the sea shore, without direct connections to the power grid through wired cables \cite{ref20}. The EH technology provides an effective approach for energy supply of maritime networks. Offshore energy sources such as wind, solar and ocean waves can be harvested to power the network equipment to support communication and computing services \cite{ref21}, \cite{ref22}. For example, ocean-wave harvested buoys anchored to the seafloor can help ensure network stability. It has been observed that one buoy is predicted to continuously generate electrical power in tens or hundreds of watts from ocean waves \cite{ref21,ref23, ref24}. An experimental research shows that the power at the level of approximately 25W to 45W can be acquired within the wave harvesting periods between 1.1s to 1.3s \cite{ref25} while the  Google’s Edge Tensor Processing Unit (TPU) computing board is capable of performing 4 trillion operations (tera-operations) per-second (TOPS) with only 2 Watts of power\cite{ref26}, \cite{ref27}. Thus, the energy generated by ocean waves may potentially meet the demands of edge computing. 
It is energy effective to deploy EH-enabled base stations with edge computing in offshore areas to fulfill the maritime communication and computing requirements, such as the processing of images/videos collected from sea lane monitoring applications. However, due to the uncertainty and variability of harvested energy, how to effectively allocate communication and computing resources and make offloading decisions for EH-enabled maritime  MEC is a difficult problem. The resource allocation must be performed by considering the utilization of harvested energy to achieve a sustainable system.

In this paper, we study the joint optimization of task scheduling and resource allocation in an EH-enabled sea lane monitoring network based on MEC. 
Since the channel states, vessel mobility, task arrivals, and energy charging and discharging vary over time, it is difficult to accurately acquire these dynamic network information. Given that the performance optimization is challenging in a highly dynamic environment, 
we consider optimizing the long-term task scheduling and resource allocation strategies to maximize the long-term average network throughput. 
The key contributions in this paper mainly include the following aspects:

\begin{itemize}
\item[$\bullet$] 
We present a two-tier MEC-based network for a sea lane monitoring scenario, which is modeled under the dual constraints on energy consumption and queue stability by considering the time-varying task arrivals, channel qualities, and available computing resources. 
\item[$\bullet$] 
To capture the dynamics of network state transitions and model the interactions between states and policies, we establish a stochastic optimization problem to maximize the time averaged network  throughput under the constraints on queue stability and energy budget. We adopt the Lyapunov optimization framework to achieve a tradeoff between network throughput and queue stability. 
\item[$\bullet$] 
Based on stochastic optimization, the formulated problem is decomposed into independent subproblems with low complexity by minimizing the upper bound of Lyapunov drift plus penalty function, which are then solved in a distributed manner, leading to an effective suboptimal solution.
\item[$\bullet$] 
We analyze the system performance and verify the asymptotic optimality of the proposed schemes. 
\end{itemize}

The rest of this paper is organized as follows. Section II provides a review of related works. The system model is presented in Section III. The problem formulation and performance evaluation are given in Sections IV and V, respectively, and the conclusion is drawn and future work is discussed in Section VI. Table I lists the key notations according to the order in which they first appear in the paper.
\begin{table}[htbp]
\centering
\caption{List of Key Notations}
\begin{tabular}{c|c}
\hline
\hline
$\mathbf{Symbol}$ & $\mathbf{Meaning}$  \\ 
\hline
$M_k$  &The numbedr of TUs under MIS $k$ \\
\hline
$\tau$  &Slot time  \\
\hline
$g_i(t)$  &The number of tasks generated at TU $i$ at time slot $t$  \\
\hline
$Y$  &Task size  \\
\hline
$g^{max}$  &Maximum number of task arrivals  \\
\hline
$\alpha$  &Number of CPU cycles for 1 bit data  \\
\hline
$T_i^{th}$  &Latency requirement of each task generated by TU $i$  \\
\hline
$\overline{g_i}$  &The average task arrival rate of TU $i$  \\ 
\hline
$y_{i,k}(t)$  &Offloading decision variable from MIS $k$ for TU $i$  \\
\hline
$z_{i,k}^n(t)$  &Subchannel allocation decision variable \\
\hline
$r_{i,k}(t)$  &The uplink achievable rate at MIS $k$ from TU $i$   \\
\hline
$p_{i,k}^n$  &Transmission power of TU $i$ on the $n$-th subchannel\\
\hline
$h_{i,k}^n(t)$  &The small-scale fading factor \\
\hline
$W$  &Subchannel bandwidth of each MIS \\ 
\hline
$\sigma^2$  &Noise power  \\
\hline
$\beta_{i,k}^n(t)$  &The largescale attenuation coefficient \\
\hline
$d_{i,k}(t)$  &the distance between TU $i$ and MIS $k$ at time slot $t$    \\
\hline
$\lambda_{k,n}$  &WaveLength of MIS $k$   \\ 
\hline
$h_i$  &The antenna heights above sea level of TU $i$   \\ 
\hline
$h_k$  &The antenna heights above sea level of MIS $k$   \\ 
\hline
$R_k$  &The radius of the optimal coverage area of MIS $k$   \\ 
\hline
$d_i^0(t)$  &The initial location TU $i$   \\ 
\hline
$v_i$  &The moving speed TU $i$   \\ 
\hline
$R_k(t)$  &The achievable transmission rate from MIS $k$ to CBS \\ 
\hline
$\rho_k$  &The radio resource allocation ratio  \\
\hline
$p_k$  &Transmission power of MIS $k$   \\
\hline
$W_c$  &Channel bandwidth of CBS \\
\hline
$\lambda_{c,k}$  &WaveLength of CBS  \\
\hline
$\beta_{k,c}^n(t)$  &The attenuation coefficient from MIS $k$ to CBS \\
\hline
$Q_i(t)$  &The length of task transmission queue of TU $i$  \\
\hline
$\theta_i(t)$  &The number of tasks offloaded in time slot $t$ \\
\hline
$F_k$  &The computing frequency of MIS $k$ \\
\hline
$f_{i,k}(t)$  &The ratio of computing resources allocated to TU $i$  \\
\hline
$Q_{i,k}(t)$  &The buffer length of MIS $k$ for TU $i$  \\
\hline
$\mu_i(t)$  &The number of tasks processed in time slot $t$ \\
\hline
$m_i(t)$  &The number of tasks migrated to CBS in time slot $t$ \\
\hline
$T_i$  &The average task latency of TU $i$ \\
\hline
$e_k(t)$  &The charging rate at MIS $k$ in time slot $t$ \\
\hline
$e_k^{max}$  &The maximum charging rate at MIS $k$ \\
\hline
$E_k(t)$  &The instantaneous energy level at MIS $k$ in time slot $t$ \\
\hline
$E_{max}$  &The maximum energy storage at MIS $k$ \\
\hline
$c_k(t)$  &The energy consumption of MIS $k$ in time slot $t$ \\
\hline
$\epsilon$  &The power coefficient of each MIS  \\
\hline
$p_{k,i}(t)$  &Processing power of MIS $k$ for TU $i$\\
\hline
\hline
\end{tabular}
\end{table}

\section{Related works}
\subsection{Maritime MEC}
As one of the key technologies in next-generation networks, MEC is considered to be leveraged in  different maritime application scenarios to support the increasing computing services. Yang \emph{et al.} \cite{ref28} mainly focused on proposing a cognitive network based on MEC for cooperative search and rescue through UAVs and unmanned ships. Distributed reinforcement learning was used to identify the channel state and perform mobile computing to optimize the data throughput in the communication group. Zeng \emph{et al.} \cite{ref29} studied the communications, computing, and caching technologies for the maritime networks based on MEC and proposed a response-based offloading algorithm to optimize task offloading. In \cite{ref30}, the authors studied a dynamic computation offloading problem to balance the tradeoff between energy and delay in offshore communication networks and proposed a two-stage joint optimal offloading algorithm (JOOA) to optimize the computation and communication resource allocation under limited energy and delay constraints. Dai \emph{et al.} \cite{ref31} focused on the unmanned aerial vehicles(UAVs) assisted multi-access computation offloading via Hybrid NOMA and FDMA in Marine Networks, aiming at minimizing the energy consumption of ocean devices.  

Existing works often focus on computation offloading or resource allocation based on one-shot optimization instead of long-term network performance maximization. 
In maritime MEC, the design of computation offloading strategies should consider the real-time changing environmental dynamics, such as time-varying channel quality and task arrivals at mobile terminals. In addition,  maritime communication and computation resources are often limited due to geographical constraints, which poses challenges for resource allocation and task offloading in maritime MEC. To improve the performance of massive ocean data transmission and task processing, it is necessary to develop a more efficient network architecture and computing solutions for maritime MEC.

\subsection{EH-enabled MEC}
Due to the technological advances in EH, off-grid renewable energy, such as solar and wind, has become promising power supply for various communication and computing networks \cite{ref32,ref33}. Most research works on EH-enabled MEC mainly focused on the EH-enabled end devices. Hu \emph{et al.} \cite{ref34} proposed an online mobility-aware offloading and resource allocation (OMORA) algorithm for EH-enabled IoT. Zhang \emph{et al.} \cite{ref35} proposed online dynamic task offloading based on the Lyapunov optimization to investigate the tradeoff between energy consumption and execution delay for a MEC system with EH. Xu \emph{et al.} \cite{ref36} first studied the resource allocation problems for EH-enabled MEC and proposed an efficient reinforcement learning (RL) based resource management algorithm to improve the system performance. Besides, Chen \emph{et al.} focused on a hybrid energy supply for powering base stations, including renewable energy and stable energy from the grid, to optimize task scheduling and energy management strategies \cite{ref37}. Mohd \emph{et al.} demonstrated that a hybrid system consisting of solar energy  and ocean waves can be an effective way to power offshore equipment\cite{ref38}.

Inspired by the aforementioned studies, we combine solar and ocean wave energies to provide hybrid energy support for maritime facilities by considering that they can compensate each other in light of the seasonal factors \cite{ref39}. In maritime scenarios, the intelligent buoys powered by hybrid solar and ocean wave energies can provide a stable energy supply for maritime MEC. However, the sustainability of green energy and the energy limitation of EH-powered devices do affect the performance of maritime MEC. 
It is non-trivial to conduct both the task scheduling and resource allocation judiciously to prevent network performance degradation caused by limited resources and network dynamics.

\subsection{Lyapunov optimization in MEC}
Lyapunov optimization has been widely applied in solving stochastic network optimization problems with “time coupling” property in task offloading and energy harvesting to avoid the prediction of dynamic variables\cite{ref40,ref41,ref42,ref43,ref44,ref45}. Applying the Lyapunov optimization, Z. Tong et al. \cite{ref40} proposed a Lyapunov online energy consumption optimization algorithm (LOECOA) to balance the system’s queue backlog and energy consumption. M. Guo et al. \cite{ref41} proposed Lyapunov-optimization-based partial computation offloading for multiuser (LOMUCO) to minimize the energy consumption of all the MDs while satisfying the constraint of time delay. Mao et al. \cite{ref42} exploited an online resource allocation approach in a multi-devices MEC system, using Gauss Seidel theory to analyze the best transmit power, but this study did not consider energy harvesting. J. Mei et al. \cite{ref43} focused on the task offloading problem for multiple energy harvesting devices and designed a maximum task offloading algorithm to maximize the system throughput based on Lyapunov optimization. However, they only simulated the performance analysis of offloading strategy and computation allocation without considering communication resource allocation. S. Bi et al. \cite{ref44} designed an online offloading scheme to combine the Lyapunov optimization and deep reinforcement learning (DRL) method together to maximize the network data processing capability. However, they just considered one MEC server which may not be applicable for multiple users. G. Ma et al. \cite{ref45} proposed a linear time complexity algorithm based on Lyapunov optimization and DRL to maximize the long-term average throughput under different constraints.

However, most existing Lyapunov optimization task offloading strategies only consider communication resource allocation or computing resource allocation during the task offloading process, and there are few studies that consider both aspects in joint optimization of task offloading, especially in EH-enabled MEC. In this paper, we leverage the Lyapunov optimization method to jointly optimize the communication and computation resources and maximize the network throughput in maritime MEC, considering the mobility of vessels, dynamic channel conditions and limited computing resources in specific marine scenarios.

\section{System model}

\subsection{Network model}
We consider a two-tier MEC-enabled sea lane monitoring network (MSLMN), as illustrated in Fig. 1. In the first network tier, a single coastal base station (CBS) centered macro-cell is deployed along the coastline to provide a wide area communication coverage for maritime information stations (MISs) near the sea lanes. In the second network tier, a set, $\mathcal{K}=\left\{1,2,...,K\right\}$, of MISs (e.g., green energy powered intelligent buoys) are anchored in advance on both sides of sea lanes within the CBS coverage area. 
Each MIS is pre-installed with solar panels and wave energy converters (WECs) to provide energy supply for communication and computing \cite{ref38}. All the MISs have non-overlapping communication coverages. Each MIS, say MIS $k$, equipped with a local MEC server, is to provide communication and computing services for terminal units (TUs) (e.g., vessels) under its coverage, the number of which is denoted by $M_k$. All the MISs are powered by solar and ocean wave energies to maintain normal operation.
The communications between CBS and MISs are based on Long Term Evolution (LTE) while WiFi is adopted for transmissions between an MIS and its associated TUs. All the TUs are randomly distributed and sailing around potentially. Each TU consistently collects the sea lane state information through captured images/videos which are further processed or computed, for the purpose of surveillance, object tracking, and monitoring.

\begin{figure}[htbp]
\centerline{\includegraphics[width=0.45\textwidth]{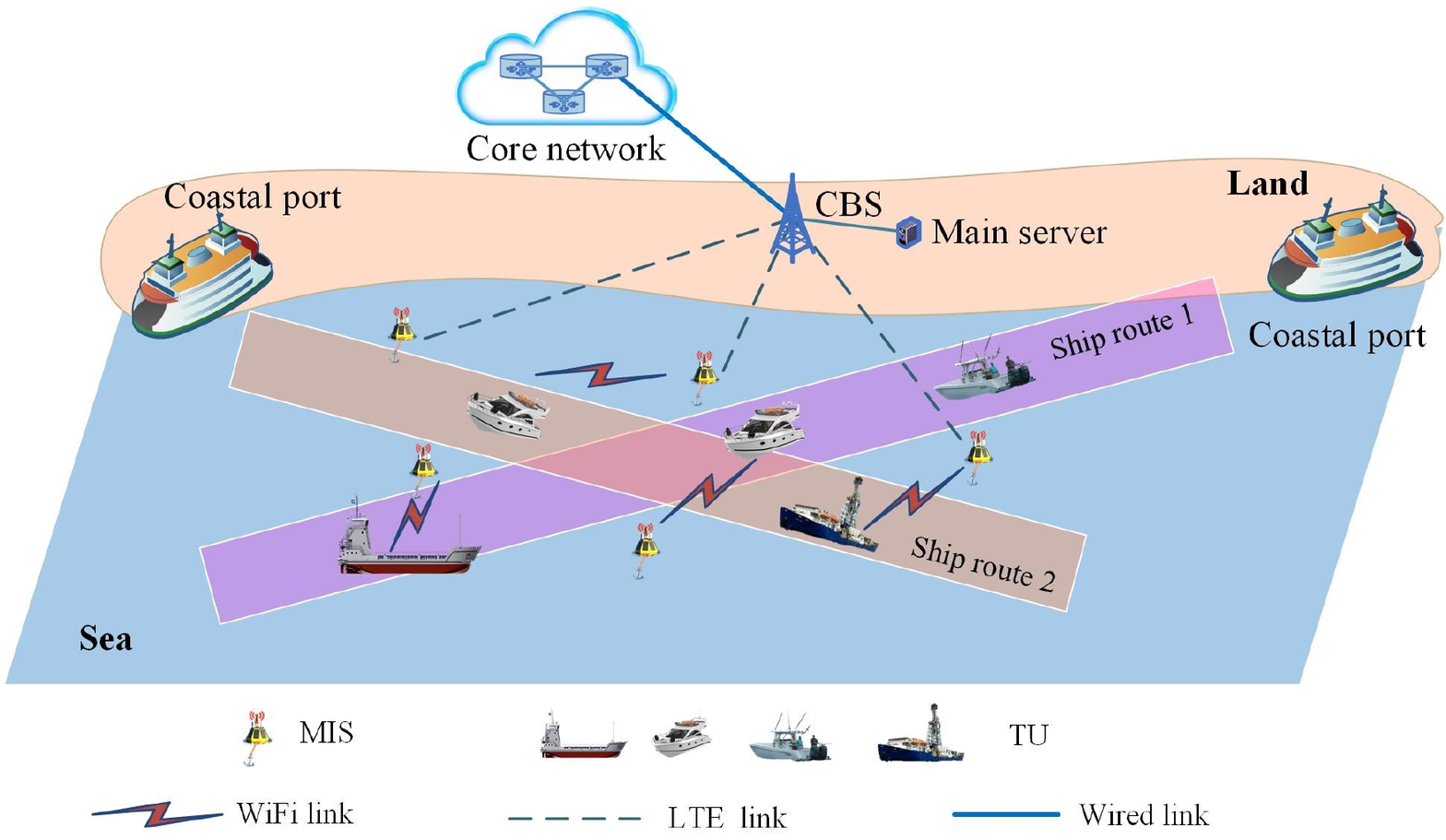}}
\caption{Network model.}
\label{fig}
\end{figure}

A two-layer edge computing infrastructure is adopted to provide near-the-TU computing capabilities. In the upper layer, a main server with abundant computation resources is connected to CBS to provide highly efficient task processing. The resources on the server are also virtualized to host different softwarized network and control functions, such as centralized radio resource control and baseband processing. Also, it can share some task processing load with different MISs through wireless transmissions especially when their energy and computation capacities are in shortage. In the lower layer, each MIS is connected with a local MEC server to provide lightweight computation and each local MEC sever hosts small-timescale virtualized network functions mainly for TUs under its coverage, e.g., task scheduling and resource management. 

We consider two timescales of system operations where the average number of TUs are assumed stationary within each large time interval (e.g., minutes to hours) for network-level radio resource reservation under the consideration of a low vessel mobility scenario. Each large time interval is discretized into a sequence of scheduling time slots and the length of each slot is denoted as $\tau$. The task offloading and resource allocation decisions are made by the MISs at the beginning of each small time slot according to the dynamic task generation and the reservation of existing resources.
We assume that each TU has limited computing capability and the generated tasks in each time slot are either offloaded to an MIS or migrated to main server connected to CBS for processing (some TUs may not be covered by CBS). The task processing results are fed back to corresponding TUs and transmitted to CBS simultaneously, which is helpful to achieve navigation efficiency and safety.


\subsection{Task arrival model}
The number of tasks generated at the $i$-th TU at time slot $t$ is assumed an independent and identically distributed random variable denoted by $g_i(t)$. In each time slot, $g_i(t)$ is randomly drawn within $ [0, g^{max}]$ to capture the temporal variations in task arrival. We assume that each task has the same size of $Y$ (in bits) and processing $1$ bit of task requires $\alpha$ CPU cycles. Additionally, each task generated by TU $i$ must be completed within its latency requirement $T_i^{th}$. Considering randomness of task arrivals, we characterize the probability density function of $g_i(t)$ at each slot $t$ as
\begin{equation}
\Pr[g_i(t)=g]=p_g
\end{equation}
where $g$ is a nonnegative integer, $g\in\mathcal{G}=\left\{0,1,2,...,g^{max}\right\}$ and $p_g\in[0,1]$. $g^{max}$ is the upper bound of the number of arriving tasks. The average task arrival rate of TU $i$ is calculated as
\begin{equation}
\overline{g_i}=\lim_{T \to \infty}\frac{1}{T}\sum_{t=0}^{T}g_i(t)=\sum_{g=0}^{g^{max}}g\cdot{p_g}.
\end{equation}

\subsection{Communication model}
Before network operation begins, CBS is preconfigured with a set of orthogonal radio spectrum resources for uplink data transmission, the amounts of which is denoted by $W_c$. We assume that each MIS has $N$ orthogonal subchannels, each with the radio bandwidth of $W$. Due to the non-overlapping communication coverages, all the MISs reuse the same portion of radio resources to exploit the resource multiplexing gain with controlled inter-MIS interference. 

At the beginning of each time slot, each MIS makes decisions on radio resource allocation among TUs under its coverage. Denote binary indicator $y_{i,k}(t)$ as the offloading decision from MIS $k$ for TU $i$ at slot $t$, which equals to $1$ when the task is offloaded to MIS $k$ and $0$, otherwise. Denote binary indicator $z_{i,k}^n(t)$ as the subchannel allocation decision, which equals to $1$ when the subchannel $n$ is allocated to TU $i$ under MIS $k$ and $0$, otherwise. According to the Shannon capacity theorem, the uplink achievable rate at MIS $k$ from TU $i$ is given by following the channel model in \cite{ref39}: 
\begin{align}
r_{i,k}(t)&=\nonumber\\
&y_{i,k}(t)\sum_{n=0}^{N}z_{i,k}^n(t)W\log_2\left(1+\frac{p_{i,k}^n\beta_{i,k}^n(t)\left| h_{i,k}^n(t) \right|^2}{\gamma+\sigma^2}\right).
\end{align}
Considering the offshore environment is relatively open with stronger direct signals and the main factors affecting the over-sea radio channels are the multi-path effect caused by sea volatility and the effect of extreme weather, we employ Rician fading channels to capture the small-scale fading characteristics of air-to-sea channels in rough sea conditions \cite{ref46, ref47}. In (3), $p_{i,k}^n$ is the transmission power at TU $i$ on the $n$-th subchannel of MIS $k$ and $h_{i,k}^n(t)$ indicates the small-scale fading factor which follows the Rician distribution with $\tilde{h}_{i,k}^n(t)=\sqrt{\frac{K_r}{1+K_r}}+\sqrt{\frac{1}{1+K_r}} s_{i,k}^n(t)$ where
$K_r$ is the Rician factor and $s_{i,k}^n(t)\in \mathcal{CN}(0,1)$\footnote{$\mathcal{CN}(0,1)$ is a complex Gaussian distribution.}. 
$\sigma^2$ is the noise power and $\gamma$ is the inter-cell interference power calculated as $\gamma=\sum_{q\in \mathcal{K}}\sum_{j\in M_q \backslash q \neq k}y_{j,q}(t)p_{j,q}^n\beta_{j,q}^n(t)$, and $\beta_{i,k}^n(t)$ is the large-scale attenuation coefficient of maritime communication expressed as \cite{ref39}
\begin{equation}
\beta_{i,k}^n(t)=\left(\frac{\lambda_{k,n}}{4\pi d_{i,k}(t)}\right)^2\left[\sin\left(\frac{2\pi h_ih_k}{\lambda_{k,n} d_{i,k}(t)}\right)\right]^2
\end{equation}
where $\lambda_{k,n}$ is the wavelength of the $n$-th subchannel of MIS $k$, $h_i$ and $h_k$  represent the antenna heights above sea level of TU $i$ and MIS $k$, respectively, and $d_{i,k}(t)$ denotes the distance between TU $i$ and MIS $k$ at the $t$-th time slot, given by
\begin{equation}
d_{i,k}(t)=
\begin{cases}
\sqrt{h_k^2+\left[\frac{R_k}{2} - d_i^0(t) - v_it\right]^2},& \text{ $\rm if$ $ d_i^0(t) \leq \frac{R_k}{2}$, } \\
\sqrt{h_k^2+\left[\frac{R_k}{2} - d_i^0(t) + v_it\right]^2},& \text{ $\rm if$ $ d_i^0(t) > \frac{R_k}{2}$,}
\end{cases}
\end{equation}
where ${R_k}$ is the radius of the optimal coverage area of MIS $k$, $d_i^0(t)$ and $v_i$ denote the initial location and moving speed of TU $i$, respectively.

When the available computing resources of an MIS is insufficient, portions of tasks are migrated to CBS through wireless transmissions for processing. The ratio of radio resource allocated to MIS $k$ from CBS, denoted as $\rho_k, 0\leq \rho_k\leq1$, is determined by CBS according to the amount of migrated tasks from different MISs. The achievable transmission rate from MIS $k$ to CBS is given by
\begin{equation}
R_k(t)=\rho_kW_c\log_2\left(1+\frac{p_k\beta_{k,c}(t)}{\sigma^2}\right)
\end{equation}
where $p_k$ is the transmission power of MIS $k$ and $\beta_{k,c}(t)$ is the attenuation coefficient from MIS $k$ to CBS.

\subsection{Communication and computation queueing models}
The length (i.e. the number of tasks) of task transmission queue of each TU at time slot $t$ is denoted by $Q_i(t)$($\leq Q_i^{\rm max}$), where $Q_i^{\rm max}$ is the buffer size of the $i$-th TU. The total number of tasks that can be offloaded in time slot $t$ is determined as $\theta_i(t)=\left\lfloor \frac{r_{i,k}(t)\cdot \tau}{Y}\right\rfloor,0\leq \theta_i(t)\leq \theta_i^{\rm max}$, and the dynamics of $Q_i(t)$ is given by
\begin{equation}
Q_i(t+1)=\max[Q_i(t)-\theta_i(t),0]+g_i(t).
\end{equation}

The computation capacity of MIS $k$ (in the unit of CPU cycles per second) is denoted as $F_k$. The ratio of computing resources allocated to TU $i$ is $f_{i,k}(t)\in[0,1]$, and we have $\mathbf{F(t)}=\left\{f_{i,k}(t)_{i\in M_k}\right\}$, where $\sum_{i=1}^{M_k}f_{i,k}(t)\leq 1$.

Suppose each MIS has a number of processing buffers for the associated TUs, which are used to accommodate offloaded tasks to be processed. The buffer length of MIS $k$  for the $i$-th TU is denoted as $Q_{i,k}(t)$($\leq Q_{i,k}^{\rm max}$), where $ Q_{i,k}^{\rm max}$ is the buffer size allocated to the $i$-th TU at MIS $k$ \cite{ref34}. The total number of tasks processed in the $t$-th time slot is expressed as  $\mu_i(t)=\left\lfloor\frac{f_{i,k}(t)F_k\cdot \tau}{\alpha Y}\right\rfloor (0\leq \mu_i(t)\leq \mu_i^{\rm max})$. The number of tasks migrated to CBS is denoted as $m_i(t)$ ($0\leq m_i(t)\leq \theta_i(t)$), and the dynamics of $Q_{i,k}(t)$ is given by
\begin{equation}
Q_{i,k}(t+1)=\max[Q_{i,k}(t)-\mu_i(t)-m_i(t),0]+\theta_i(t).
\end{equation}

The average queue length of the $i$-th TU is given by
\begin{equation}
Q_i^{ava}=\lim_{T \to \infty}\frac{1}{T}\sum_{t=0}^{T}\left\{Q_i(t)+Q_{i,k}(t)\right\}.
\end{equation}


According to Little’s Law, the average queuing latency is proportional to the average queue length \cite{ref35}, i.e., $T_i^{ava}=\frac{Q_i^{ava}}{\overline{g_i}}$, where $\overline{g_i}$ denotes the task arrival rate. Furthermore, the average task latency of TU $i$  mainly consists of the average queuing delay and the average task execution delay (i.e., a constant $T^c$)\cite{ref48}\cite{ref49}, given by
\begin{equation}
T_i=\frac{Q_i^{ava}}{\overline{g_i}}+T^c.
\end{equation}
In this work, we mainly focused on uplink computation offloading and communication resource allocation and assume that the downlink transmission time can be neglected considering each computing result is usually small in size, which is also a common  assumption made in \cite{ref6}, \cite{ref48} and \cite{ref49}.

\subsection{Energy model}
Considering that in a long run (e.g., over consecutive days) the energy collection behavior can follow a periodic pattern, we model the dynamic energy changing process over small-timescale scheduling slots as a stochastic process to reflect the dynamics of energy input, where the changing rate at each slot fluctuates due to the intermittent energy sources, such as the intensity of solar radiation or wave heights in different geographic locations.
We assume each MIS is equipped with an energy queue with finite capacity for storing the renewable energy generated from ocean waves and solar energy. The energy charging rate of each MIS changes over time depending on the features of energy sources, such as the intensity of solar radiation or wave heights in different geographic locations. We model the energy charging process at each MIS as a stochastic process where the charging rate at MIS $k$ in the $t$-th time slot is assumed to be an independent random variable $e_k(t)$, uniformly distributed within the interval of $[0, e_k^{\rm max}]$ \cite{ref50}. The maximum energy storage of each MIS is denoted by $E_{\rm max}$, then the instantaneous energy level at the $t$-th time slot is represented by $E_k(t)$, which is updated over consecutive time slots, given by \cite{ref50} 
\begin{equation}
E_k(t+1)=\min[\max(0,E_k(t)+e_k(t)-c_k(t)), E_{\rm max}]
\end{equation}
where $E_k(0)=0$. $c_k(t)$ is the energy consumption of the MIS at the $t$-th time slot, for equipment maintenance, task transmission, and computation, i.e., given by
\begin{equation}
c_k(t)=c_k^{bas}(t)+c_k^{tra}(t)+c_k^{com}(t).
\end{equation}

We use a widely adopted power consumption model for computing  $p_{k,i}(t)=\epsilon[f_{i,k}(t)F_k]^3$ \cite{ref28}, where $\epsilon$ is a constant power coefficient which depends on the chip structure of MIS. Then the computation energy consumption is given by
\begin{equation}
c_k^{com}(t)=\sum_{i\in M_k}p_{k,i}(t)\tau=\sum_{i\in M_k}\epsilon[f_{i,k}(t)F_k]^3\tau.
\end{equation}

As CBS has high computation capacity, the processing delay for each migrated task at an MIS is too small to be negligible \cite{ref51}. Then, the transmission energy consumption mainly refers to the energy consumed by task migration from an MIS to CBS given by
\begin{equation}
c_k^{tra}(t)=p_k\cdot \frac{\sum_{i\in M_k}m_i(t)Y}{R_k(t)}.
\end{equation}

\section{Problem formulation}
We consider the average network throughput as the optimization objective, which is calculated as the expected aggregate transmission rate of the whole system in each time slot, given by
\begin{equation}
\mathcal{H}(t)=\sum_{k\in \mathcal{K}}\sum_{i\in M_k}r_{i,k}(t),
\end{equation}
and the statistical average of $\mathcal{H}(t)$ over $T$ time slots is given by
\begin{equation}
\overline{\mathcal{H}(t)}=\lim_{T \to \infty}\frac{1}{T}\sum_{t=0}^{T-1}\mathbb{E}\left\{ \mathcal{H}(t) \right\}.
\end{equation}

Mathematically, the problem is formulated as:
\begin{displaymath}
(\mathbf{P1}):\max_{y_{i,k}(t),z_{i,k}^n(t),f_{i,k}(t),m_i(t)}\overline{\mathcal{H}(t)},
\end{displaymath}
\begin{align}
s.t.\nonumber\\
&{\rm(C1)}:0 \leq c_k(t)\leq E_k(t), \forall k\in \mathcal{K},t, \nonumber\\
&{\rm(C2)}:T_i\leq T_i^{th}, \forall i,\nonumber\\
&{\rm(C3)}:\sum_{i\in M_k}f_{i,k}(t)\leq 1, \forall k,t,\nonumber\\
&{\rm(C4)}:f_{i,k}(t)\in[0,1], \forall i,k,t,\nonumber\\
&{\rm(C5)}:y_{i,k}(t)\in\left\{ 0,1 \right\}, \forall i,k,t,\nonumber\\
&{\rm(C6)}:\sum_{i\in M_k}z_{i,k}^n(t)\leq 1, \forall k,n,t,\nonumber\\
&{\rm(C7)}:z_{i,k}^n(t)\in\left\{ 0,1 \right\}, \forall i,k,n,t,\nonumber\\
&{\rm(C8)}:0\leq \rho_k\leq1, \forall k,\nonumber\\
&{\rm(C9)}:0\leq \theta_i(t)\leq\theta_i^{\rm max}, \forall i,t,\nonumber\\
&{\rm(C10)}:0\leq \mu_i(t)\leq\mu_i^{\rm max}, \forall i,t,\nonumber\\
&{\rm(C11)}:0\leq m_i(t)\leq \theta_i(t)-\mu_i(t), \forall i,t.
\end{align} 

In (17), (C1) and (C2) are the energy and average latency constraints. (C3) implies that the per-slot computation load on each MIS cannot exceed its capacity. (C6) indicates that a subchannel on MIS $k$ can be occupied by at most one TU in each time slot. (C11) means that the number of migrated tasks of each TU under MIS $k$ cannot exceed the number of arrivals in each time slot.

Problem ($\mathbf{P1}$) is a stochastic optimization problem with large state and action spaces, where the per-time-slot problem is also a mixed-integer non-linear program (MINLP) since it involves binary variables determining task offloading and communication/computation resource allocation. The major challenge in deriving the optimal solution of ($\mathbf{P1}$) is the lack of future information (e.g., task arrivals and dynamic channel states), which varies over time and is difficult to predict in advance. In addition, the queue stability constraint (C2) incurs the coupling of system variables over time. We leverage the Lyapunov optimization framework to address this challenge by decomposing ($\mathbf{P1}$) into a series of deterministic optimization problems at each time slot and obtaining the asymptotic optimal solutions at the premise of ensuring the system stability\cite{ref52}.

In ($\mathbf{P1}$), the energy constraint (C1) is a typical temporal coupling problem, which couples the peer task offloading decisions across different time slots. Since (C1) is to ensure that the energy of MIS $k$ at the $t$-th time slot meets the energy consumption requirements in the time slot, we construct a virtual energy queue (denoted as $Z_k(t)$) for (C1) and transform it into a queue stability problem by employing Lyapunov optimization method. The dynamics of the virtual queue is given by
\begin{equation}
Z_k(t+1)=\max[Z_k(t)+c_k(t)-E_k(t),0].
\end{equation}

Obviously, the queue length cannot keep increasing with time if the energy consumption requirements is satisfied. Then, ($\mathbf{P1}$) can be transformed to jointly stabilize the queue and maxmize the average network throughput, which make it more tractable.

\begin{theorem} 
If $\lim_{T \to \infty}\frac{\mathbb{E}\left\{ Z_k(t) \right\}}{t}=0$, the virtual queue is stable with the constraint, $c_k(t)\leq E_k(t)$, being satisfied. 
\end{theorem}

The proof of Theorem 1 is provided in APPENDIX A.

Let $\mathbf{\Theta}(t)=[Q_i(t),Q_{i,k}(t),Z_k(t)]$. Define the Lyapunov function as
\begin{equation}
\mathbf{L}(\mathbf{\Theta}(t))=\frac{1}{2}\sum_{k\in \mathcal{K}}\left\{ Z_k^2(t)+\sum_{i\in M_k}[Q_i^2(t)+Q_{i,k}^2(t)] \right\}.
\end{equation}


Then, the \emph{Lyapunov drift} function can be used to push the Lyapunov function to a stable state and maintain stability of all queues \cite{ref53}, i.e., to make the queues stable by minimizing the upper bound of the \emph{Lyapunov drift} function, which is defined as
\begin{align}
\Delta(\mathbf{\Theta}(t))&= \mathbb{E}\left\{\mathbf{L}(\mathbf{\Theta}(t+1))-\mathbf{L}(\mathbf{\Theta}(t)) \vert \mathbf{\Theta}(t)\right\}\nonumber\\
     &\underset{(*)}{\leq} C+\sum_{k\in \mathcal{K}} Z_k(t)[c_k(t)-E_k(t)]\nonumber\\
     &+ \sum_{k\in \mathcal{K}}\sum_{i\in M_k}Q_i(t)[g_i(t)-\theta_i(t)]\nonumber\\
     &+ \sum_{k\in \mathcal{K}}\sum_{i\in M_k}Q_{i,k}(t)[\theta_i(t)-\mu_i(t)-m_i(t)],
\end{align}
where $C$ is a constant. The proof of inequality (*) is given in APPENDIX B.

The objective of ($\mathbf{P1}$) is to maximize the total throughput of the system. When $r_{i,k}(t)$ increases, the transmitted tasks and system energy consumption will also increase due to (8) and (12), which in turn increases the queue length. Therefore, the two objectives of maintaining queue stability and maximizing the average network throughput are incompatible and cannot be optimized at the same time. An alternative solution to this problem is to merge the two objectives into a single objective with a utility function, i.e., \emph{drift plus penalty function}, which is expressed as
\begin{displaymath}
\Gamma(\mathbf{\Theta}(t))=\Delta(\mathbf{\Theta}(t))-V\cdot\mathbb{E}\left\{ \mathcal{H}(t)\vert \mathbf{\Theta}(t) \right\}
\end{displaymath}
\begin{displaymath}
\leq C-V\mathbb{E}\left\{ \mathcal{H}(t)\vert \mathbf{\Theta}(t) \right\}+\sum_{k\in \mathcal{K}}Z_k(t)\mathbb{E}[c_k(t)-E_k(t)]
\end{displaymath}
\begin{displaymath}
+\sum_{k\in \mathcal{K}}\sum_{i\in M_k}Q_i(t)\mathbb{E}[g_i(t)-\theta_i(t)]
\end{displaymath}
\begin{equation}
+\sum_{k\in \mathcal{K}}\sum_{i\in M_k}Q_{i,k}(t)\mathbb{E}[\theta_i(t)-\mu_i(t)-m_i(t)]
\end{equation}
where $\Gamma(\mathbf{\Theta}(t))$ indicates the average drift of the Lyapunov function value and the tradeoff between throughput and queue length in consecutive time slots, which is used to measure the stability of the system. 
The parameter $V$ is a non-negative constant used to control the weight between queue length and throughput. By adjusting the value of $V$, the trade-off between queue stability and time average throughput can be achieved.

In this way, the long-term queue stability constraint and the long-term network throughput can be integrated into one optimization objective. Then the original problem $\mathbf{P1}$ is transformed into a new optimization problem which only relies on the information of current time slot and next time slot. 
To further eliminate the dependence on the next time slot information, we found the upper bound of $\Gamma(\mathbf{\Theta}(t))$ and shifted towards minimizing the upper bound instead of minimizing $\Gamma(\mathbf{\Theta}(t))$.
Then, our optimization objective is transformed into minimizing the upper bound of  $\Gamma(\mathbf{\Theta}(t))$ in each time slot. According to the concept of opportunistically minimizing an expectation, minimizing $f(x)$ can ensure that $\mathbb{E}[f(x)]$ is minimized \cite{ref54}, we minimize the supremum of $\Gamma(\mathbf{\Theta}(t))$ in ($\mathbf{P2}$) by removing the conditional expectation of (21).
\begin{displaymath}
(\mathbf{P2}):\min_{y_{i,k}(t),z_{i,k}^n(t),f_{i,k}(t),m_i(t)} \sum_{k\in \mathcal{K}}Z_k(t)[c_k(t)-E_k(t)]
\end{displaymath}
\begin{displaymath}
-V\sum_{k\in \mathcal{K}}\sum_{i\in M_k}r_{i,k}(t)+\sum_{k\in \mathcal{K}}\sum_{i\in M_k}[Q_{i,k}(t)-Q_i(t)]\theta_i(t)
\end{displaymath}
\begin{displaymath}
+\sum_{k\in \mathcal{K}}\sum_{i\in M_k}[Q_i(t)g_i(t)-Q_{i,k}(t)\mu_i(t)-Q_{i,k}(t)m_i(t)],
\end{displaymath}
\begin{equation}
{\rm s.t.\ (C1)-(C11)}.
\end{equation}

In (22), the terms $Q_i(t)g_i(t)$ and $Z_k(t)E_k(t)$ are independent of the decision variables and hence can be ignored in this optimization problem. We decompose ($\mathbf{P2}$) into ($\mathbf{P2.1}$) and ($\mathbf{P2.2}$) with the consideration of task transmission and task processing independently. In ($\mathbf{P2.1}$), $y_{i,k}(t)$ and $z_{i,k}^n(t)$ are task offloading and subchannel allocation decision variables, respectively. In ($\mathbf{P2.2}$), $f_{i,k}(t)$ and $m_i(t)$ are task migration and computing resource allocation decision variables, respectively.
\begin{displaymath}
(\mathbf{P2.1}):\min_{y_{i,k}(t),z_{i,k}^n(t)} \sum_{k\in \mathcal{K}}\sum_{i\in M_k} \left\{[Q_{i,k}(t)-Q_i(t)]\theta_i(t)\right.
\end{displaymath}
\begin{displaymath}
\left.-Vr_{i,k}(t)\right\},
\end{displaymath}
\begin{equation}
{\rm s.t.\ (C5),(C6),(C7),(C9)}.
\end{equation}


\subsection{Task offloading}

By substituting $\theta_i(t)$ with $\left\lfloor\frac{r_{i,k}(t)\cdot \tau}{Y}\right\rfloor$ in ($\mathbf{P2.1}$), we have
\begin{displaymath}
\min_{y_{i,k}(t),z_{i,k}^n(t)} \sum_{k\in \mathcal{K}}\sum_{i\in M_k} \left\{[Q_{i,k}(t)-Q_i(t)]\left\lfloor\frac{r_{i,k}(t)\cdot \tau}{Y}\right\rfloor\right.
\end{displaymath}
\begin{displaymath}
\left.-Vr_{i,k}(t)\right\},
\end{displaymath}
\begin{equation}
{\rm s.t.\ (C5),(C6),(C7),(C9)}.
\end{equation}

For brevity, we let 
\begin{equation}
G(t)=\sum_{n=0}^{N}z_{i,k}^n(t)W\log_2\left(1+\frac{p_{i,k}^n\beta_{i,k}^n(t)}{\gamma+\sigma^2}\right), 
\end{equation}
and then we have
\begin{equation}
r_{i,k}(t)=y_{i,k}(t)G(t).
\end{equation}

By substituting (26) into (24), we transform ($\mathbf{P2.1}$) into
\begin{displaymath}
\min_{y_{i,k}(t)} \sum_{k\in \mathcal{K}}\sum_{i\in M_k} \left\{[Q_{i,k}(t)-Q_i(t)]\right.
\end{displaymath}
\begin{displaymath}
\left.\left\lfloor\frac{y_{i,k}(t)G(t)\cdot \tau}{Y}-V y_{i,k}(t)G(t)\right\rfloor\right\},
\end{displaymath}
\begin{equation}
{\rm s.t.\ (C5)}.
\end{equation}

We observe that (27) is a linear function of $y_{i,k}(t)$, then we obtain the optimal solution of the offloading decision as
\begin{equation}
y_{i,k}^*(t)=
\begin{cases}
1,& \text{ $\frac{[Q_{i,k}(t)-Q_i(t)]G(t)\cdot \tau}{Y}-VG(t)\leq 0$, } \\
0,& \text{ $\rm otherwise.$ }
\end{cases}
\end{equation}

\subsection{Subchannel allocation}

When $y_{i,k}(t)$ is determined, we rewrite (24) as
\begin{displaymath}
\min_{z_{i,k}^n(t)} \sum_{k\in \mathcal{K}}\sum_{i\in M_k}\sum_{n=0}^{N} \left\{\frac{[Q_{i,k}(t)-Q_i(t)]\cdot \tau}{Y}-V\right\}\cdot
\end{displaymath}
\begin{displaymath}
z_{i,k}^n(t)W\log_2\left(1+\frac{p_{i,k}\beta_{i,k}^n(t)}{\gamma+\sigma^2}\right),
\end{displaymath}
\begin{equation}
{\rm s.t.\ (C6),(C7)}.
\end{equation}

For all $k\in \mathcal{K}$, $i\in M_k$ and $n \in N$, we let 
\begin{displaymath}
 W_{i,k}^n(t)= \left\{\frac{[Q_{i,k}(t)-Q_i(t)]\cdot \tau}{Y}-V\right\}\cdot W\log_2
\end{displaymath}
\begin{equation}
\left(1+\frac{p_{i,k}^n\beta_{i,k}^n(t)}{\sum_{q\in \mathcal{K}}\sum_{j\in M_q(t) \backslash q \neq k}y_{j,q}(t)p_{j,q}^n(t)\beta_{j,q}^n(t)+\sigma^2}\right),
\end{equation}
which represents the weight of TU $i$ on subchannel $n$.Then, for the $n$-th subchannel, we get the subchannel allocation indicator $z_{i,k}^n(t)$ as:
\begin{equation}
z_{i,k}^n(t)=
\begin{cases}
1,& \text{ $ i=\arg\min W_{i,k}^n(t)$, } \\
0,& \text{ $\rm otherwise.$ }
\end{cases}
\end{equation}

Based on this observation, we design a subchannel allocation scheme to meet the communication requirements of different TUs, as shown in Algorithm 1. We use $A_{i,k}(t)$ to represent the set of subchannels allocated to each TU, and $B_k(t)$ to denote the remaining available subchannel resources of MIS $k$. 



\begin{table}[!htbp] 
\centerline{ 
\linespread{1.25}

\small

\begin{tabular}{p{8cm} p{2.5cm}{l}}

\toprule 

\textbf{Algorithm 1} subchannel allocation algorithm. \\  

\midrule 

1: \textbf{Input:}\\ 

\quad {At the beginning of each time slot t, obtain $\beta_{i,k}^n(t)$,}\\
\quad{$A_{i,k}(t)$ and $B_k(t)$}.\\

2:\quad\emph{\textbf{while} $B_k(t)\neq \varnothing$ \textbf{do}}\\

3:\quad\quad\emph{\textbf{for} int $k = 0,k \leq M_k,k++$ \textbf{do}}\\

4:\quad\quad\quad{\emph{Calculate $W_{i,k}^n(t)$ with formula $(30)$.}}\\

5:\quad\quad\emph{\textbf{endfor}}\\

6:\quad\quad{\emph{obtain the optimal $i^*$ from formula $(31)$.}}\\

7:\quad\quad{\emph{let $z_{i^*,k}^n{^*}(t)=1$ if $ i=arg min W_{i,k}^n(t)$.}}\\

8:\quad\quad{\emph{update:$A_{i,k}(t)=A_{i,k}(t)\cup \left\{n^*\right\}$, }}\\
  \quad\quad\quad\quad\quad{\emph{ $B_k(t) =B_k(t)-\left\{n^*\right\}$.}}\\

9:\quad\emph{\textbf{endwhile}}\\

\bottomrule 

\end{tabular}
} 

\end{table} 

After solving $(\mathbf{P2.1})$, we attempt to find the optimal solution to $(\mathbf{P2.2})$, which is expressed as 

$(\mathbf{P2.2}):$
\begin{displaymath}
\min_{f_{i,k}(t),m_i(t)}\left\{Z_k(t)c_k(t)-\sum_{i\in M_k}Q_{i,k}(t)[\mu_i(t)+m_i(t)]\right\},
\end{displaymath}

\begin{equation}
{\rm s.t.\ (C3),(C4),(C8),(C10),(C11)}.
\end{equation}

By substituting $\mu_i(t)$ and $c_k(t)$ with $\mu_i(t)=\left\lfloor\frac{f_{i,k}(t)F_k\cdot \tau}{\alpha Y}\right\rfloor$ and (12), we transform ($\mathbf{P2.2}$) into
\begin{displaymath}
\min_{f_{i,k}(t),m_i(t)}Z_k(t)\left\{c_k^{bas}(t)+p_k\frac{\sum_{i\in M_k}m_i(t)Y}{R_k(t)}\right.
\end{displaymath}
\begin{displaymath}
\left.+\sum_{i\in M_k}\epsilon(f_{i,k}(t)F_k)^3\tau\right\}
\end{displaymath}
\begin{displaymath}
-\sum_{i\in M_k}\left\{Q_{i,k}(t)\left\lfloor\frac{f_{i,k}(t)F_k\cdot \tau}{\alpha Y}\right\rfloor+Q_{i,k}(t)m_i(t)\right\}
\end{displaymath}
\begin{displaymath}
=\min_{f_{i,k}(t),m_i(t)}\sum_{i\in M_k}\left\{\frac{Z_k(t)p_kY}{R_k(t)}m_i(t)-Q_{i,k}(t)m_i(t)\right\}
\end{displaymath}
\begin{displaymath}
+\sum_{i\in M_k}\left\{Z_k(t)\epsilon(f_{i,k}(t)F_k)^3\tau-Q_{i,k}(t)\left\lfloor\frac{f_{i,k}(t)F_k\cdot \tau}{\alpha Y}\right\rfloor\right\}
\end{displaymath}
\begin{displaymath}
+Z_k(t)c_k^{bas}(t),
\end{displaymath}
\begin{equation}
{\rm s.t.\ (C3),(C4),(C8),(C10),(C11)}.
\end{equation}

In (33), $Z_k(t)c_k^{bas}(t)$ is independent of decision variable in this optimization and hence can be ignored. The remaining items of (33) can be divided into two major problems, i.e., task migration and computing resource allocation, respectively.

\subsection{Task migration}
\begin{displaymath}
\min_{m_i(t)} \sum_{i\in M_k}\left\{\frac{Z_k(t)p_kY}{R_k(t)}m_i(t)-Q_{i,k}(t)m_i(t)\right\},
\end{displaymath}
\begin{equation}
{\rm s.t.\ (C8), (C11)}.
\end{equation}
We observe that (34) is a linear program which can be optimized at each time slot. The optimal solution is
\begin{equation}
m_i^*(t)=
\begin{cases}
\theta_i(t)-\mu_i(t),& \text{ $\frac{Z_k(t)p_kY}{R_k(t)}-Q_{i,k}(t)\leq 0$ } \\
0,& \text{ $\rm otherwise.$ }
\end{cases}
\end{equation}

When $\frac{Z_k(t)p_kY}{R_k(t)}-Q_{i,k}(t)\leq 0$, the task buffer of TU $i$ at MIS $k$ is large and the channel condition from MIS $k$ to CBS is better with larger $R_k(t)$. In this case, the best way is to migrate all the tasks to CBS for computing. Otherwise, it is preferable to process all the tasks  from TU $i$ at MIS $k$ when the queue length $Q_{i,k}(t)$ is short.

\subsection{Computing resource allocation}
\begin{displaymath}
\min_{f_{i,k}(t)} \sum_{i\in M_k}\left\{Z_k(t)\epsilon(f_{i,k}(t)F_k)^3\tau-Q_{i,k}(t)\left\lfloor\frac{f_{i,k}(t)F_k}{\alpha Y}\right\rfloor\right\},
\end{displaymath}
\begin{equation}
{\rm s.t.\ (C3),(C4)}.
\end{equation}

The objective function in (36) is convex which can be proved by the second-order derivative. Since each $f_{i,k}(t)$ is mutually independent under one MIS, we can easily obtain the optimal value for each TU under (C4). However, sometimes if all the TUs take the optimal value of $f_{i,k}(t)$, (C3) cannot be satisfied. In this case, we evenly allocate the computing resources under the MIS. Then, the optimal computing resource allocation variable $f_{i,k}(t)$ for MIS $k$ to process the computing task of TU $i$ is given by
\begin{equation}
f_{i,k}^*(t)=
\begin{cases}
\sqrt{\frac{Q_{i,k}(t)}{3\alpha YZ_k(t)\epsilon F_k^2}},& \text{ $\sum_{i\in M_k}f_{i,k}(t) \leq 1,$ } \\
\frac{\sqrt{Q_{i,k}(t)}}{\sum_{i\in M_k}\sqrt{Q_{i,k}(t)}},& \text{ $\sum_{i\in M_k}f_{i,k}(t) > 1.$ }
\end{cases}
\end{equation}

On this basis, we propose a computing resource allocation algorithm to solve (36), which is described in Algorithm 2.
\begin{table}[!htbp] 
\centerline{ 

\linespread{1.25}

\small

\begin{tabular}{p{8cm} p{2.5cm}{l}}

\toprule 

\textbf{Algorithm 2} Computational resource allocation algorithm. \\  

\midrule 

1: \textbf{Input:}{ $\alpha$, $Y$, $\tau$, $\epsilon$. }\\ 

2: \emph{\textbf{Initialization:} $t\leftarrow 0,$}\\
  \quad\emph{$Q_k(0)=0,Q_{i,k}(0)=0,Z_k(0)=0$.}\\
3:\quad\emph{\textbf{while} $t\leq T$ \textbf{do}}\\

4:\quad\quad\emph{At the beginning of each time slot $t$, obtain $Q_k(t),$} \\
  \quad\quad\quad\emph{$Q_{i,k}(t),Z_k(t)$.}\\

5:\quad\quad{\emph{Determine $f_{i,k}^*(t)$ with formula $(36)$ and $(37)$.}}\\

6:\quad\emph{\textbf{endwhile}}\\


\bottomrule 

\end{tabular}
} 

\end{table} 

By optimizing task offloading, subchannel resource  allocation, task migration and computing resource allocation, we propose a \underline{j}oint \underline{c}omputation \underline{o}ffloading and \underline{r}esource \underline{a}llocation algorithm (JCORA) for the optimization objective. The whole process is described in Algorithm 3.

\begin{table}[!htbp] 
\centerline{ 

\linespread{1.25}

\small

\begin{tabular}{p{8cm} p{2.5cm}{l}}

\toprule 

\textbf{Algorithm 3} JCORA: Joint computation offloading and resource allocation algorithm. \\  

\midrule 

1: \emph{\textbf{Initialization:}$t\leftarrow 0,$}\\
   \quad\emph{$Q_k(0)=0,Q_{i,k}(0)=0,Z_k(0)=0$.}\\

2:\quad\emph{\textbf{while} $t\leq T$ \textbf{do}}\\

3:\quad\quad\emph{At the beginning of each time slot $t$, obtain $Q_k(t),$} \\
  \quad\quad\quad\emph{$Q_{i,k}(t),Z_k(t)$.}\\

4:\quad\quad{\emph{Obtain $z_{i,k}^n(t)$ by calling Algorithm 1.}}\\

5:\quad\quad{\emph{Determine $y_{i,k}(t)$ with formula $(28)$.}}\\

6:\quad\quad{\emph{Calculate $\theta_i(t)$ by $\theta_i(t)=\left\lfloor \frac{r_{i,k}(t)\cdot \tau}{Y}\right\rfloor$.}}\\

7:\quad\quad{\emph{Calculate $(34)$ and get $ m_i(t)$ with $(35)$.}}\\

8:\quad\quad{\emph{Run Algorithm 2 to determine the computational}}\\
   \quad\quad\quad{resource allocation policy $f_{i,k}^*(t)$.}\\

9:\quad\emph{\textbf{endwhile}}\\

10: \quad\emph{Update the queues and $t\leftarrow t+1$.}\\

\bottomrule 

\end{tabular}
} 

\end{table} 

Next, we rigorously analyze the performance of JCORA theoretically. Specifically, we demonstrate that the average network throughput achieved by JCORA can arbitrarily approximate the optimal solution and that there exists a supremum bound for the expected average queue length, which are summarized in Theorem 2. 

\begin{theorem} 
For any $V>0$, the performance gap between the optimal solution of the original problem $(\mathbf{P1})$ (denoted as $\overline{\mathcal{H}^*(t)}$) and the result of JCORA is represented by
\begin{equation}
\overline{\mathcal{H}^*(t)} - \overline{\mathcal{H}(t)}\leq \frac{C}{V}
\end{equation}
and the average queue length is upper bounded by
\begin{equation}
\overline{Q}\leq \frac{1}{\xi}\left\{C+V\cdot[{\mathcal{H}(t)}^{max}-{\mathcal{H}(t)}^{min}]\right\},
\end{equation}
where ${\mathcal{H}(t)}^{max}$ and ${\mathcal{H}(t)}^{min}$ represent the maximum and minimum values of the original problem, respectively, $\xi$ is a real number greater than $0$.
\end{theorem}

The proof of Theorem 2 is provided in APPENDIX C.

Theorem 2 sheds light on an $[O(\frac{1}{V}), O(V)]$ tradeoff between average network throughput and queue length (or, latency).  Formula (38) demonstrates that the gap between the network throughput of JCORA and the optimal solution is upper bounded by $\frac{C}{V}$. If $V$ is sufficiently large, the average network throughput of JCORA can be asymptotically close to the optimum value. In (39), the average queue length of JCORA is proved to be upper bounded, which illustrates that JCORA can always maintain the stability of all the queues. Intuitively, the performance of JCORA depends on the control parameter $V$, which can be tuned flexibly to improve the network throughput, but at the price of a larger delay since the average queue length increases linearly with $V$. Thus, by adjusting the control parameter $V$, we can dexterously reap the balance between the average network throughput and queue length.

Then, we give the computational complexity of the proposed JCORA algorithm which mainly consists of five parts, i.e., lines 2–8 in Algorithm 3. Since the task offloading decision is made by each MIS, its computation overhead is linear with the number of MIS s, i.e., $O(KM_k)$. In Algorithm 1, we first obtain the computation overhead of subchannel allocation of which is $O(NM_k)$. Thus, the computation overhead of subchannel allocation is $O(KNM_k)$. The Task migration is calculated with (35) and the computation overhead of which is liner with the number of TUs, i.e., $O(KM_k)$. Similar to the subchannel allocation, the computation overhead of computation resource allocation at one MIS is also $O(M_k)$ since each MIS calculates the CPU-cycle frequency $f_{i,k}(t)$ via (37). Therefore, the computational complexity of the proposed JCORA algorithm is $O(KM_k) + O(NM_k) + O(KNM_k) + O(KM_k) = O(KNM_k)$.

\section{Performance Evaluation}
In this section, we evaluate the performance of the proposed JCORA algorithm through simulations, which is done under different parameters of $V$, and compared with other three resource allocation schemes in terms of average throughput and average latency. As we mentioned earlier, We consider the average network throughput as the optimization objective, which is calculated as the expected aggregate transmission rate of the whole system in each time slot.
\subsection{Simulation setup}
All the simulations are conducted using MATLAB on a PC configured with a Core i7-10510U 1.80 GHz CPU and 8 GB of RAM. We consider an offshore network consisting of 1 CBS and 5 MISs which are deployed in a $400\times400$ $m^2$ area, together with 5-30 random sailing TUs. In this system, MIS is responsible for making resource allocation decisions for each TU, so as to maximize the network throughput under the constraints of average latency and power consumption. 
The overall spectrum bandwidth of CBS is 100 MHz and each MIS has 30 subchannels for computation offloading of its associated TUs. The  bandwidth of each subchannel is $1$ MHz. The power spectral density of the additive white Gaussian noise is $-174$ dBm/Hz. The antenna heights above sea level of TU $i$ and MIS $k$ are $10m$ and $50m$ respectively. We assume each MIS communicates with the CBS over LTE and with its associated TUs over WiFi. All the TUs are randomly distributed and sailing around potentially. We assume all the MISs reuse the same portion of radio resources to exploit the resource multiplexing gain with controlled inter-MIS interference. Other simulation parameters are shown in Table II.

\begin{table}[htbp]
\centering
\caption{Simulation Parameter Settings}
\begin{tabular}{c|c}
\hline
\hline
Parameters &Values  \\ 
\hline
Subchannel bandwidth, $W$  &$1MHz$  \\ 
\hline
Noise power, $\sigma^2$  &$-174dBm/Hz$  \\
\hline
Transmission power of TU, $p_{i,k}^n$  &$0.1W$  \\ 
\hline
Transmission power of MIS $k$, $p_k$  &$1W$  \\
\hline
The maximum energy storage of each MIS, $E_{max}$   &$20J$  \\
\hline
The computing frequency of MIS, $F_k$  &$1G$ $cycles/s$  \\
\hline
Slot time, $\tau$  &$0.05s$  \\
\hline
The power coefficient of each MIS,$\epsilon$  &$1e-25$  \\
\hline
WaveLength of MIS $k$, $\lambda_{k,n}$  &$0.125m$  \\
\hline
WaveLength of CBS, $\lambda_{c,k}$  &$0.02m$  \\
\hline
Number of CPU cycles for 1 bit data, $\alpha$  &$1000 cycles/bit$  \\
\hline
Task size, $Y$  &$1000bit$  \\
\hline
The radio resource allocation ratio, $\rho_k$  &$0.1$  \\
\hline
Maximum number of task arrivals, $g^{max}$  &$300$  \\
\hline
\hline
\end{tabular}
\end{table}

\subsection{Performance evaluation}
\emph{1) Throughput-latency Tradeoff:} Fig. 2 illustrates the effect of control parameter $V$ on the performance of JCORA algorithm. As $V$ increases, the average network throughput becomes high, which is consistent with our theoretical analysis. For a larger $V$, the algorithm emphasizes on the network  throughput more than on the queue stability, and thus the maximum network capacity can be reached when $V$ approaches to 1. JCORA dynamically tunes resource allocation decisions to improve the average throughput. In addition, the latency increases with control parameter $V$, because a large $V$ implies a heavy weight on achieving high throughput, which leads to the increase of average queue length and queuing delay. From Fig. 2, we can see that the proposed algorithm balances throughput with latency by adjusting the value of $V$. 
\begin{figure}[htbp]
\centerline{\includegraphics[width=0.50\textwidth]{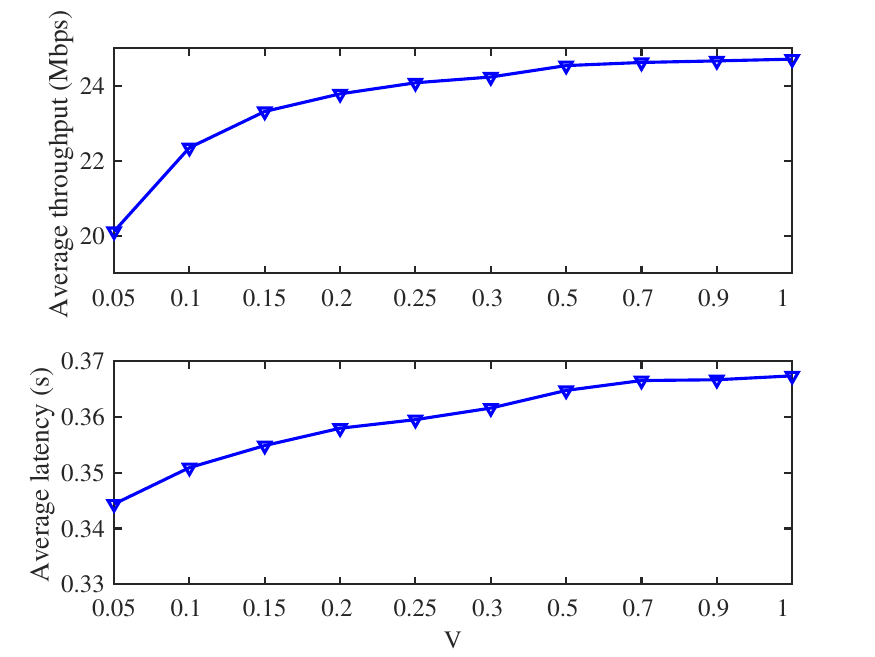}}
\caption{The average throughput and average latency versus $V$.}
\label{fig}
\end{figure}


\emph{2) Energy Queue Length and Energy Consumption:} Fig. 3 shows the change of energy queue length under different control parameter $V$. We observe that the energy queue length tends to decrease with the increase of $V$. This is because when $V$ increases, the average throughput of the system becomes larger, resulting in the increase of system energy consumption and the decrease of energy queue length. Fig. 4 illustrates the average energy consumption and the corresponding energy queue length with respect to different number of time slots. The results show that the average energy consumption is always lower than the energy queue length in different time slots, which satisfies the constraints (C1) in (17). In this way, the long-term energy constraints of MISs is satisfied.
\begin{figure}[htbp]
\centerline{\includegraphics[width=0.45\textwidth]{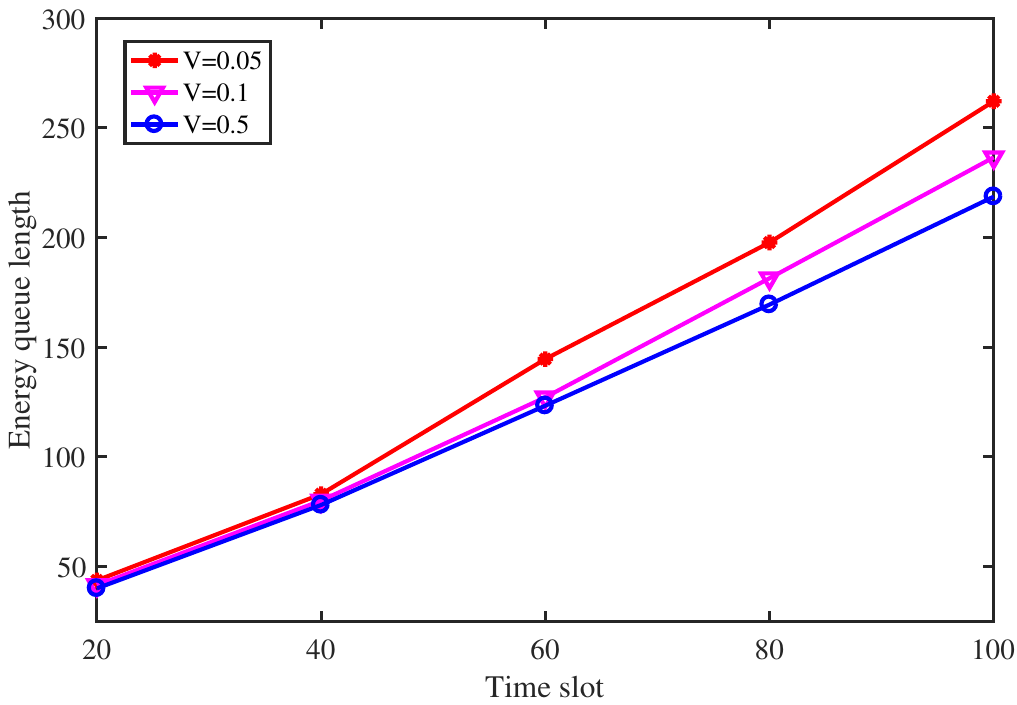}}
\caption{The energy queue length versus time slot.}
\label{fig}
\end{figure}

\begin{figure}[htbp]
\centerline{\includegraphics[width=0.45\textwidth]{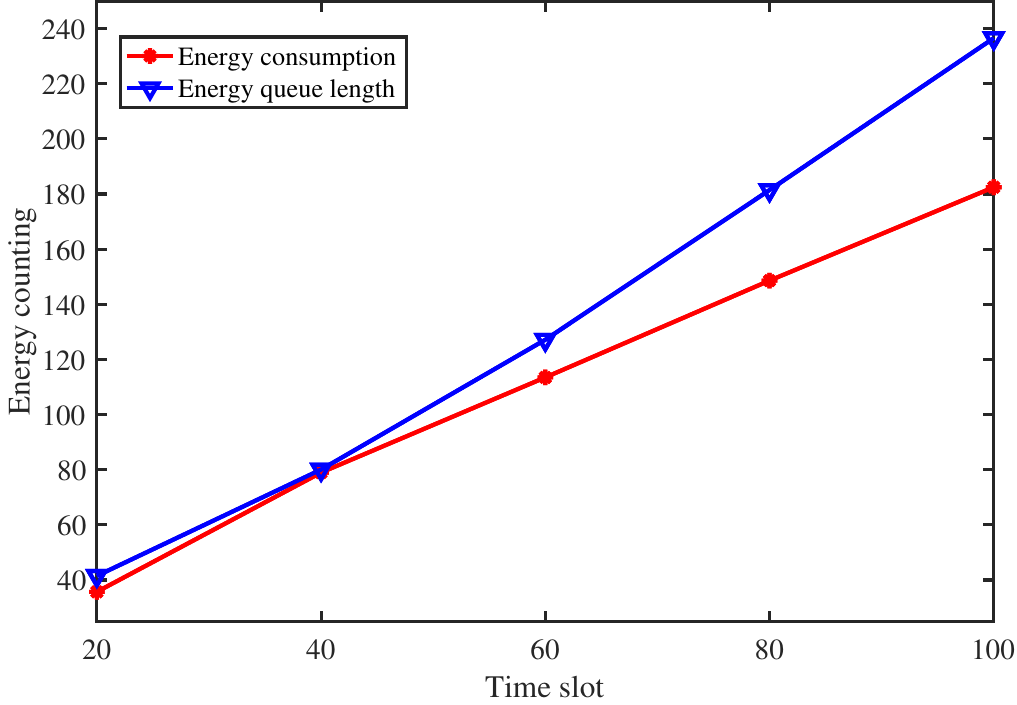}}
\caption{The average energy consumption versus energy queue length}
\label{fig}
\end{figure}




\emph{3) Effect of the Number of TUs:} Fig. 5 and Fig. 6 illustrate the average throughput and average latency of the system with respect to different number of TUs under different values of control parameter $V$. When the number of TUs increases, the average throughput is improved at the expense of increased latency. As seen in Fig. 5 and Fig. 6, JCORA can adapt to different number of TUs and improve the average network throughput at a tolerable latency.
\begin{figure}[htbp]
\centerline{\includegraphics[width=0.45\textwidth]{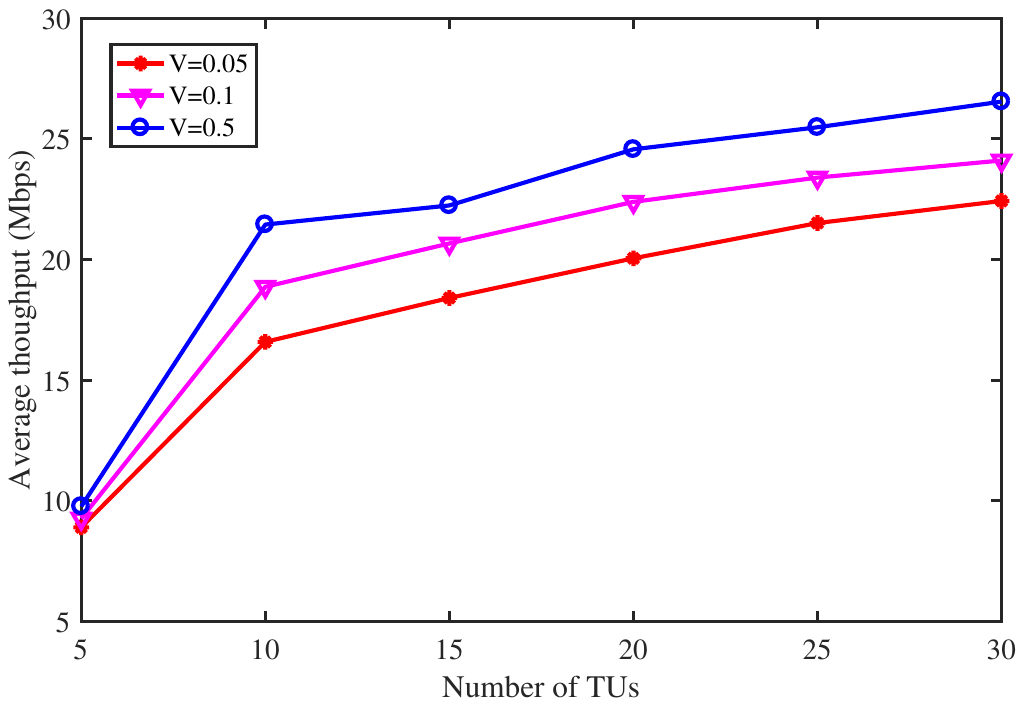}}
\caption{The average throughput versus number of TUs}
\label{fig}
\end{figure}

\begin{figure}[htbp]
\centerline{\includegraphics[width=0.45\textwidth]{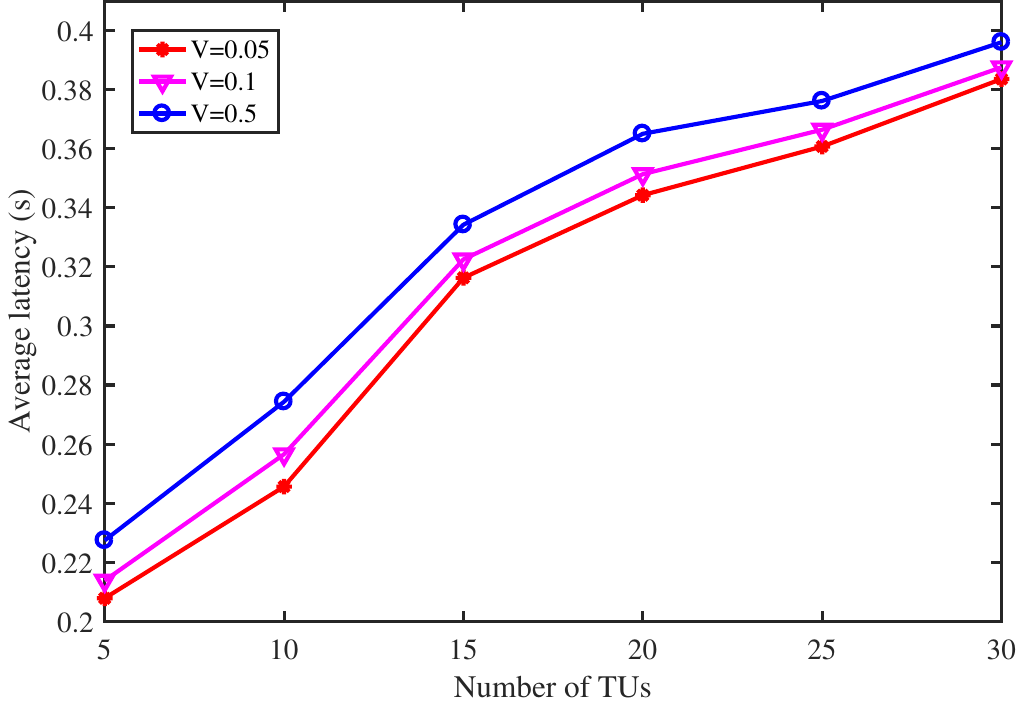}}
\caption{The average latency versus number of TUs}
\label{fig}
\end{figure}

\subsection{Performance comparison}
We further compare the performance of JCORA ($V$ is set to 0.1) with four benchmark algorithms listed as follows.
\begin{itemize}
\item[$a)$] 
First-in-first-out (FIFO) based resource allocation algorithm (FRA) \cite{ref30}: The communication and computation resources of MIS $k$ are allocated to the first packet arrival sequence of TUs, and the other TUs wait in line following the M/M/1 queuing process.
\item[$b)$] 
Latency based resource allocation algorithm (LRA): Under the energy limitation of MIS in each time slot, $f_{i,k}(t)$ is determined by the latency constraint of each TU under MIS $k$. If $f_{i,k}(t)$ exceeds the maximum computing frequency that MIS $k$ can provide, the optimal resource allocation solution is the maximum $f_{i,k}(t)$ for each TU under the maximum tolerance time of TU $i$ \cite{ref55}.
\item[$c)$] 
Priority based resource allocation algorithm (PRA): Each TU under MIS $k$ is prioritized by the task arrival rate $g_i(t)$. The higher the task arrival rate, the higher the priority obtained, and the more advantageous the resource allocation, then the communication and computation resources are allocated according to the obtained priority \cite{ref56}.  
\item[$d)$] 
TDMA (time division multiple access) based resource allocation algorithm (TRA): Multiple TUs can access a subchannel under one MIS at different times in a time slot\cite{ref57}.  
\end{itemize}

Fig. 7 demonstrates the average throughput with respect to the number of TUs for five different algorithms. We observe that the average throughput of JCORA is the highest compared with FRA, LRA, PRA and TRA. The reason is that JCORA can adjust the decisions of task offloading and resource allocation according to the changes of channel state, real-time energy state and random task arrivals, etc. In contrast, LRA, PRA and TRA do not consider the network dynamics when making resource allocation decisions while FRA allocate resources to only one TU at a time. Fig. 8 shows a comparison of average latency with respect to the number of TUs achieved by different algorithms. We observe that the latency of JCORA is lower than that of FRA, LRA, PRA and TRA. The latency of FRA is the largest, because FRA can only process tasks for one TU at a time slot. The more the TUs, the longer the queueing delay. In Fig. 7 and Fig. 8, we observe the advantages of JCORA in improving the average network throughput and stabilizing the queue length. 

\begin{figure}[htbp]
\centerline{\includegraphics[width=0.45\textwidth]{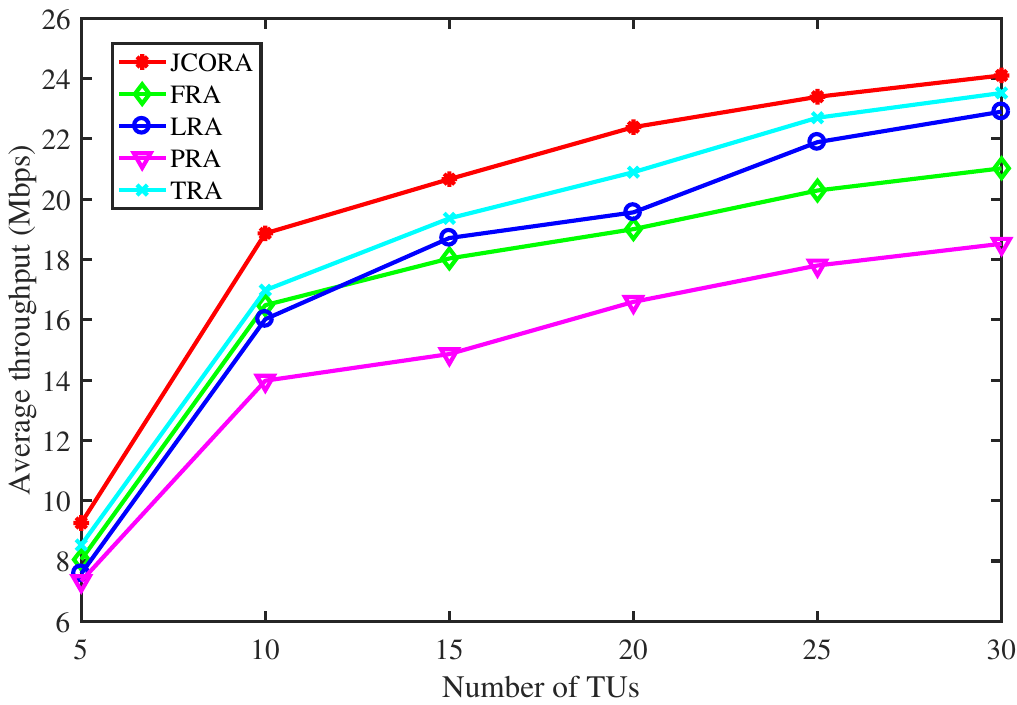}}
\caption{The average throughput versus number of TUs.}
\label{fig}
\end{figure}

\begin{figure}[htbp]
\centerline{\includegraphics[width=0.45\textwidth]{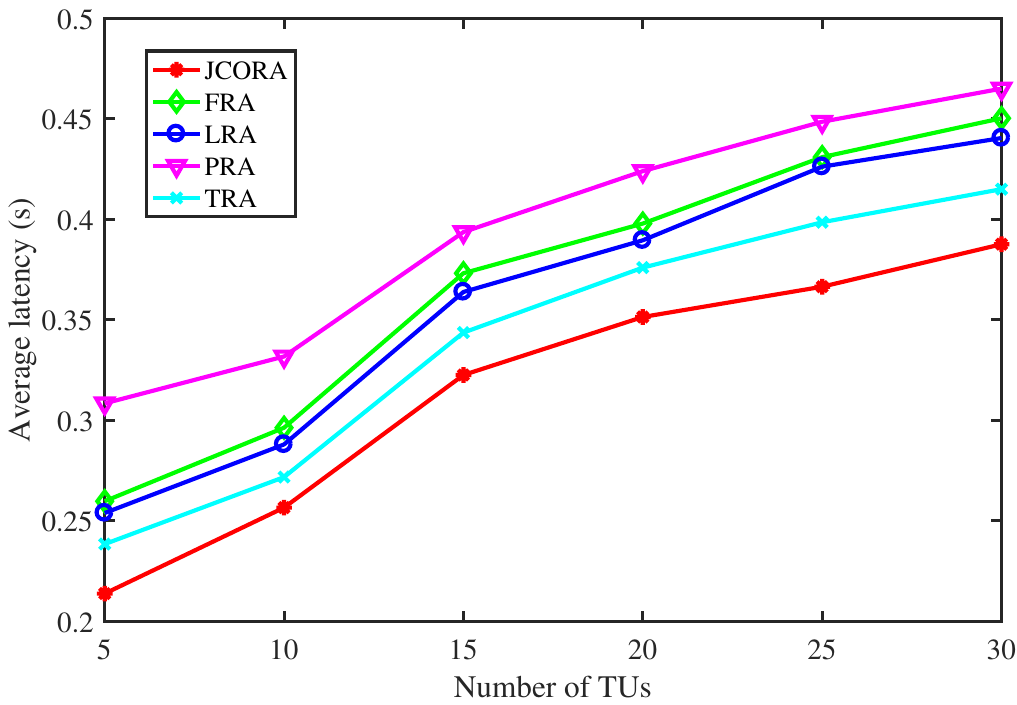}}
\caption{The average latency versus number of TUs.}
\label{fig}
\end{figure}

We further compare the average throughput and average latency of the five algorithms under different task arrival rates in Fig. 9 and Fig. 10. We observe that both the average throughput and average latency are in a positive correlation with task arrival rate. Among them the average throughput of JCORA outperforms the others, because JCORA makes resource allocation and computation offloading decisions by adapting to the dynamic network environment. The performance of TRA is also superior to the other three algorithms, as the adopt of TDMA can also achieve better relatively allocation of communication resources.

\begin{figure}[htbp]
\centerline{\includegraphics[width=0.45\textwidth]{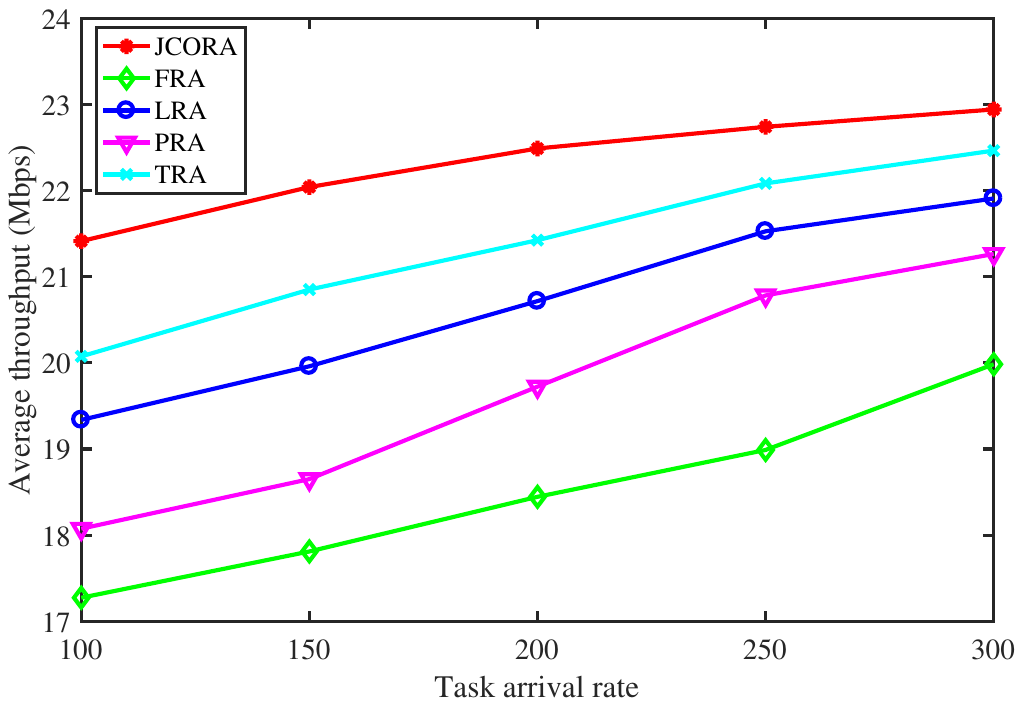}}
\caption{The average throughput versus task arrival rate.}
\label{fig}
\end{figure}

\begin{figure}[htbp]
\centerline{\includegraphics[width=0.45\textwidth]{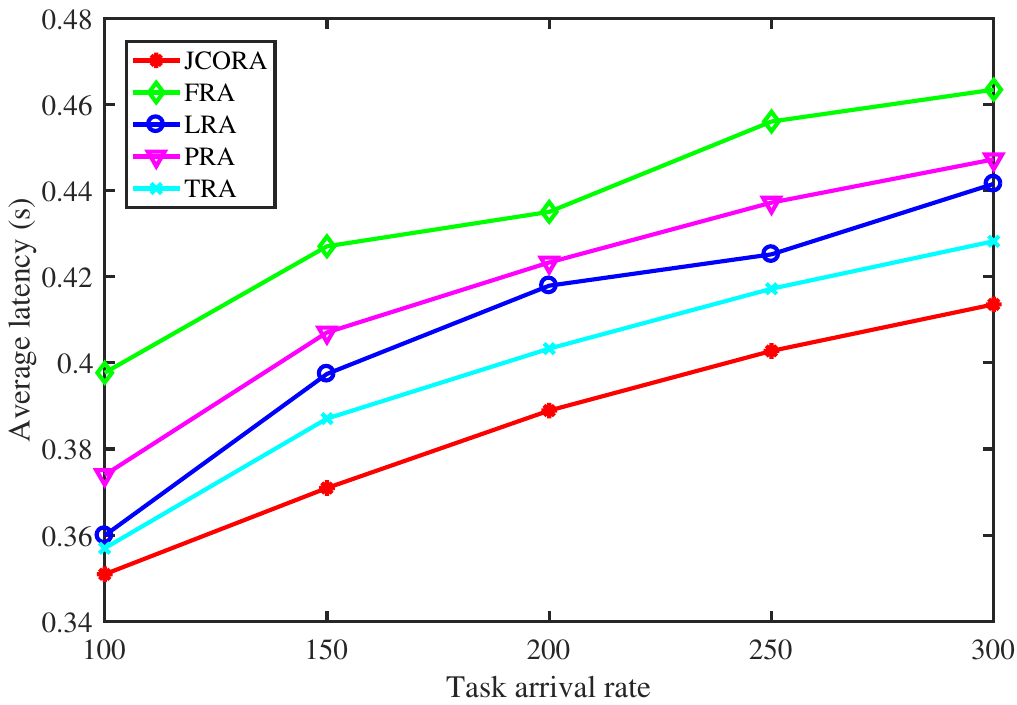}}
\caption{The average latency versus task arrival rate.}
\label{fig}
\end{figure}

Finally, we compare the average throughput of the five algorithms under different maximal energy charging rates in Fig. 11 and Fig. 12. We observe that the performance of JCORA outperforms the others, and the average throughput increases when the maximal energy charging rate is larger while  the latency is just the opposite, due to the fact that the greater the energy obtained by the MIS, the more tasks can be processed, and the processing speed will also be improved.

\begin{figure}[htbp]
\centerline{\includegraphics[width=0.45\textwidth]{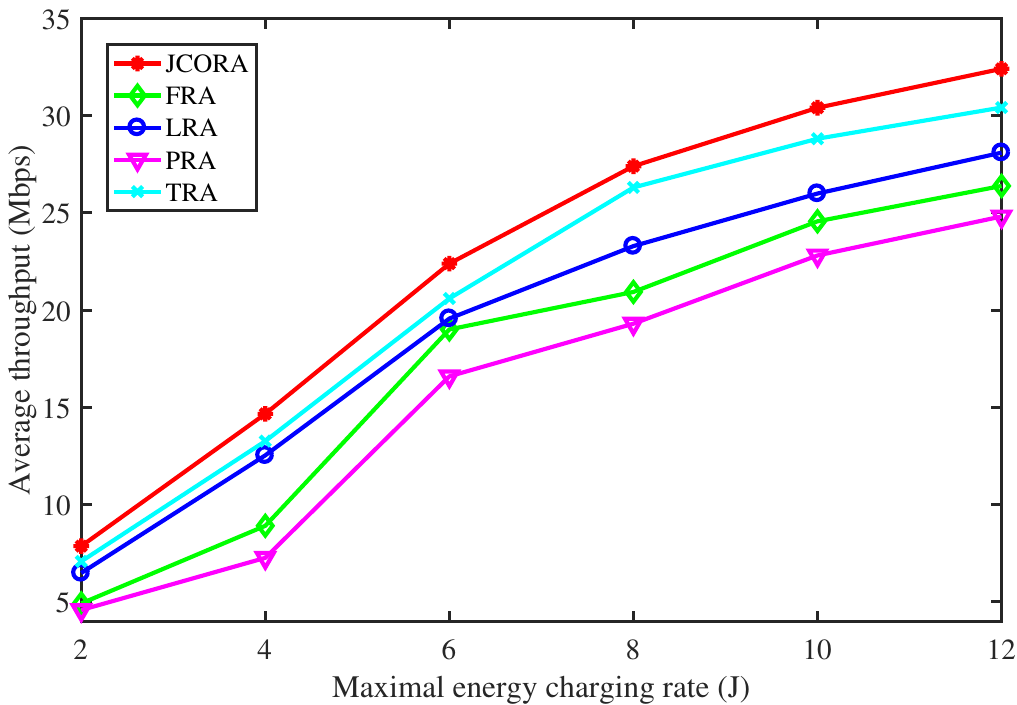}}
\caption{The average throughput versus maximal energy charging rate.}
\label{fig}
\end{figure}

\begin{figure}[htbp]
\centerline{\includegraphics[width=0.45\textwidth]{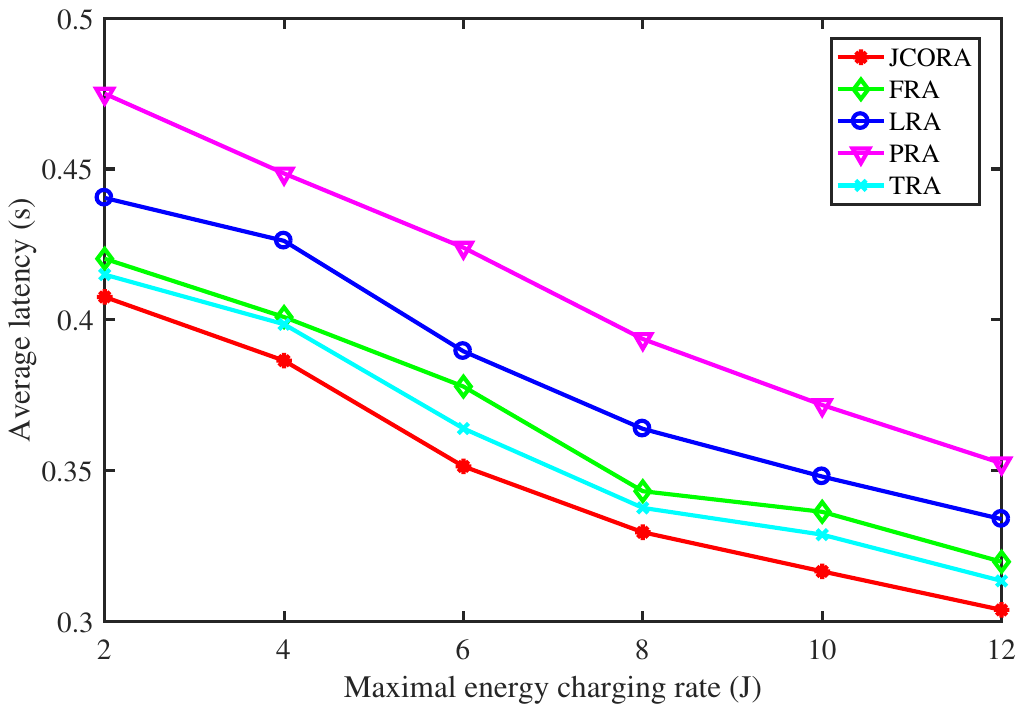}}
\caption{The average latency versus maximal energy charging rate.}
\label{fig}
\end{figure}

\section{Conclusion and Future Work}
In this paper, we have considered a sea lane monitoring network with MEC and EH and investigated a throughput-queue stability tradeoff for dynamic computation offloading. With stable task and energy queues, we have proposed a JCORA algorithm based on the Lyapunov optimization to obtain the joint computation offloading and resource allocation decisions. We transform the original problem into a deterministic program, which is decoupled into multiple independent subproblems for optimization. The performance analysis is carried out to reveal the asymptotic optimality of the proposed algorithms and demonstrate the superiority over other benchmark schemes. Our study provides a feasible approach to design future offshore MEC-enabled networks with renewable energy powered edge servers. For future work, we will study resource allocation and computation offloading problems for more complex and dynamic marine scenarios (e.g., environment monitoring) by exploring machine learning techniques (e.g., deep reinforcement learning) for decision making.



%

\appendices
\section{Proof of Theorem 1}
$\mathbf{Proof}$:
\begin{displaymath}
Z_k(t+1)=\max[Z_k(t)+c_k(t)-E_k(t),0] 
\end{displaymath}
\begin{equation}
\geq Z_k(t)+c_k(t)-E_k(t),
\end{equation}
\begin{equation}
\frac{Z_k(t)}{t}\geq \frac{Z_k(0)}{t}+\frac{1}{t}\sum_{s=0}^{t-1}[c_k(x)-E_k(x)],
\end{equation}
\begin{equation}
\frac{\mathbb{E}\left\{Z_k(t)\right\}-Z_k(0)}{t}\geq \frac{1}{t}\sum_{s=0}^{t-1}\mathbb{E}\left\{c_k(x)-E_k(x)\right\},
\end{equation}
\begin{equation}
\lim_{t \to \infty}\frac{\mathbb{E}\left\{Z_k(t)\right\}}{t}\geq \lim_{t \to \infty}\frac{1}{t}\sum_{s=0}^{t-1}\mathbb{E}\left\{c_k(x)-E_k(x)\right\}.
\end{equation}

If $Z_k(t)$ is stable, $\lim_{T \to \infty}\frac{\mathbb{E}\left\{ Z_k(t) \right\}}{t}=0$, which means that the constraint($c_k(t)\leq E_k(t)$) is satisfied. 

\section{Proof of (*) in (20)}

$\mathbf{Proof}$:
\begin{displaymath}
\mathbf{L}(\mathbf{\Theta}(t+1))-\mathbf{L}(\mathbf{\Theta}(t))=\frac{1}{2}\sum_{k\in \mathcal{K}}[Z_k^2(t+1)-Z_k^2(t)]
\end{displaymath}
\begin{displaymath}
+\frac{1}{2}\sum_{k\in \mathcal{K}}\sum_{i\in M_k}[Q_i^2(t+1)-Q_i^2(t)]
\end{displaymath}
\begin{equation}
+\frac{1}{2}\sum_{k\in \mathcal{K}}\sum_{i\in M_k}[Q_{i,k}^2(t+1)-Q_{i,k}^2(t)].
\end{equation}

Based on the inequality
\begin{equation}
[\max(X-Y,0)+A]^2\leq X^2+Y^2+A^2+2X(A-Y),
\end{equation}
we can obtain
\begin{displaymath}
Z_k^2(t+1)-Z_k^2(t)=[Z_k(t)+c_k(t)-E_k(t)]^2-[Z_k(t)]^2
\end{displaymath}
\begin{displaymath}
= [c_k(t)]^2+[E_k(t)]^2+2Z_k(t)[c_k(t)-E_k(t)]-2c_k(t)E_k(t)
\end{displaymath}
\begin{equation}
\leq [c_k(t)]^2+[E_k(t)]^2+2Z_k(t)[c_k(t)-E_k(t)].
\end{equation}

Similarly, we have
\begin{displaymath}
Q_i^2(t+1)-Q_i^2(t)
\end{displaymath}
\begin{equation}
\leq [\theta_i(t)]^2+[g_i(t)]^2+2Q_i(t)[g_i(t)-\theta_i(t)]
\end{equation}
and
\begin{displaymath}
Q_{i,k}^2(t+1)-Q_{i,k}^2(t)\leq [\theta_i(t)]^2+[\mu_i(t)]^2+[m_i(t)]^2
\end{displaymath}
\begin{equation}
+2Q_{i,k}(t)[\theta_i(t)-\mu_i(t)-m_i(t)].
\end{equation}

Combining (46), (47), (48), we have
\begin{displaymath}
\mathbf{L}(\mathbf{\Theta}(t+1))-\mathbf{L}(\mathbf{\Theta}(t))\leq C+\sum_{k\in \mathcal{K}} Z_k(t)[c_k(t)-E_k(t)]
\end{displaymath}
\begin{displaymath}
+\sum_{k\in \mathcal{K}}\sum_{i\in M_k}Q_i(t)[g_i(t)-\theta_i(t)]
\end{displaymath}
\begin{equation}
+ \sum_{k\in \mathcal{K}}\sum_{i\in M_k}Q_{i,k}(t)[\theta_i(t)-\mu_i(t)-m_i(t)],
\end{equation}
where $C$ can be described as:
\begin{displaymath}
C=\frac{1}{2}\sum_{k\in \mathcal{K}}c_k(t)]^2+[E_k(t)]^2+\sum_{k\in \mathcal{K}}\sum_{i\in M_k}[\theta_i(t)]^2+[g_i(t)]^2]
\end{displaymath}
\begin{displaymath}
+\sum_{k\in \mathcal{K}}\sum_{i\in M_k}[\theta_i(t)]^2+[\mu_i(t)]^2+[m_i(t)]^2]\leq
\end{displaymath}
\begin{equation}
 \sum_{k\in \mathcal{K}}\left\{ E_{max}^2+\sum_{i\in M_k}[(\theta_{i}^{max})^2+(g_{i})^{max})^2+\frac{1}{2}(\mu_{i}^{max})^2] \right\}.
\end{equation}

\section{Proof of Theorem 2}
$\mathbf{Proof}$:

According to Caratheodory’s theorem \cite{ref53}, there always exists an optimal policy $\pi_1$, which satisfies:
\begin{displaymath}
\mathbb{E}\left\{\mathcal{H}(t) \vert\pi_1 \right\}= \overline{\mathcal{H}^*(t)},
\end{displaymath}
\begin{displaymath}
\mathbb{E}\left\{c_k(t)\vert\pi_1 \right\}\leq \mathbb{E}\left\{E_k(t)\vert\pi_1 \right\},
\end{displaymath}
\begin{displaymath}
\mathbb{E}\left\{g_i(t)\vert\pi_1 \right\}\leq \mathbb{E}\left\{\theta_i(t)\vert\pi_1 \right\},
\end{displaymath}
\begin{equation}
 \mathbb{E}\left\{\theta_i(t)\vert\pi_1 \right\}\leq \mathbb{E}\left\{\mu_i(t)+m_i(t)\vert\pi_1 \right\}.
\end{equation}

By substituting (51) into (21), we have
\begin{displaymath}
\Delta(\mathbf{\Theta}(t))-V\cdot\mathbb{E}\left\{ \mathcal{H}(t)\vert \mathbf{\Theta}(t) \right\}
\end{displaymath}
\begin{displaymath}
\leq C-V\overline{\mathcal{H}^*(t)} + \sum_{k\in \mathcal{K}}Z_k(t)\mathbb{E}\left\{[c_k(t)-E_k(t)]\vert\pi_1\right\}
\end{displaymath}
\begin{displaymath}
+\sum_{k\in \mathcal{K}}\sum_{i\in M_k}Q_i(t)\mathbb{E}\left\{[g_i(t)-\theta_i(t)]\vert\pi_1\right\}
\end{displaymath}
\begin{displaymath}
+\sum_{k\in \mathcal{K}}\sum_{i\in M_k}Q_{i,k}(t)\mathbb{E}\left\{[\theta_i(t)-\mu_i(t)-m_i(t)]\vert\pi_1\right\}
\end{displaymath}
\begin{equation}
\leq C-V\overline{\mathcal{H}^*(t)} + 0.
\end{equation}

Then for a stable system, we have
\begin{equation}
\Delta(\mathbf{\Theta}(t))-V\cdot\mathbb{E}\left\{ \mathcal{H}(t)\vert \mathbf{\Theta}(t) \right\}\leq C-V\overline{\mathcal{H}^*(t)}
\end{equation}
and 
\begin{equation}
\sum_{t=0}^{T-1}\Delta(\mathbf{\Theta}(t))=\mathbf{L}(\mathbf{\Theta}(T)) < \infty.
\end{equation}

Combining (53) and (54), we have
\begin{displaymath}
\lim_{T \to \infty}\frac{1}{T}\sum_{t=0}^{T-1}\Delta(\mathbf{\Theta}(t))-V\cdot\lim_{T \to \infty}\frac{1}{T}\sum_{t=0}^{T-1}\mathbb{E}\left\{ \mathcal{H}(t) \right\}
\end{displaymath}
\begin{equation}
= 0 - V\cdot\lim_{T \to \infty}\frac{1}{T}\sum_{t=0}^{T-1}\mathbb{E}\left\{ \mathcal{H}(t) \right\}\leq C-V\overline{\mathcal{H}^*(t)}.
\end{equation}

Dividing (55) by $V$, we can obtain
\begin{equation}
\overline{\mathcal{H}^*(t)} - \overline{\mathcal{H}(t)}\leq \frac{C}{V}
\end{equation}
where $\overline{\mathcal{H}(t)}=\lim_{T \to \infty}\frac{1}{T}\sum_{t=0}^{T-1}\mathbb{E}\left\{ \mathcal{H}(t) \right\}$.

We assume that for TU $i \in M_k$ , there exists some real number $\xi > 0$ under policy $\pi_2$, satisfying
\begin{displaymath}
\mathbb{E}\left\{\mathcal{H}(t) \vert\pi_2 \right\}= \mathcal{H}^\xi(t),
\end{displaymath}
\begin{displaymath}
\mathbb{E}\left\{[c_k(t)-E_k(t)]\vert\pi_2 \right\}\leq -\xi,
\end{displaymath}
\begin{displaymath}
\mathbb{E}\left\{[g_i(t)-\theta_i(t)]\vert\pi_2 \right\}\leq -\xi,
\end{displaymath}
\begin{equation}
\mathbb{E}\left\{[\theta_i(t)-\mu_i(t)-m_i(t)]\vert\pi_2 \right\}\leq -\xi.
\end{equation}

According to (52), we have
\begin{displaymath}
\Delta(\mathbf{\Theta}(t))-V\cdot\mathbb{E}\left\{ \mathcal{H}(t)\vert \mathbf{\Theta}(t) \right\}
\end{displaymath}
\begin{displaymath}
\leq C-V\mathcal{H}^\xi(t) + \mathbb{E}\sum_{k\in \mathcal{K}}Z_k(t)\left\{[c_k(t)-E_k(t)]\vert\pi\right\}
\end{displaymath}
\begin{displaymath}
+\mathbb{E}\sum_{k\in \mathcal{K}}\sum_{i\in M_k}Q_i(t)\left\{[g_i(t)-\theta_i(t)]\vert\pi\right\}
\end{displaymath}
\begin{displaymath}
+\mathbb{E}\sum_{k\in \mathcal{K}}\sum_{i\in M_k}Q_{i,k}(t)\left\{[\theta_i(t)-\mu_i(t)-m_i(t)]\vert\pi\right\}\leq 
\end{displaymath}
\begin{equation}
C-V\mathcal{H}^\xi(t) -\xi\mathbb{E}\left\{\sum_{k\in \mathcal{K}}Z_k(t)+\sum_{k\in \mathcal{K}}\sum_{i\in M_k}[Q_i(t)+Q_{i,k}(t)]\right\}.
\end{equation}

Combining (54) and (58), we have
\begin{displaymath}
\lim_{T \to \infty}\frac{1}{T}\sum_{s=0}^{T-1}\Delta(\mathbf{\Theta}(t))-V\cdot\lim_{T \to \infty}\frac{1}{T}\sum_{t=0}^{T-1}\mathbb{E}\left\{ \mathcal{H}(t) \right\}
\end{displaymath}
\begin{displaymath}
= 0 - V\cdot\overline{\mathcal{H}(t)} \leq C-V\mathcal{H}^\xi(t)
\end{displaymath}
\begin{equation}
-\xi\lim_{T \to \infty}\frac{1}{T}\sum_{t=0}^{T-1}\left\{\sum_{k\in \mathcal{K}}Z_k(t)+\sum_{k\in \mathcal{K}}\sum_{i\in M_k}[Q_i(t)+Q_{i,k}(t)]\right\}.
\end{equation}

Dividing (59) by $\xi$, we can obtain
\begin{displaymath}
\overline{Q}\leq \frac{1}{\xi}\left\{C+V\cdot[\overline{\mathcal{H}(t)}-\mathcal{H}^\xi(t)]\right\}
\end{displaymath}
\begin{equation}
\leq \frac{1}{\xi}\left\{C+V\cdot[{\mathcal{H}(t)}^{max}-{\mathcal{H}(t)}^{min}]\right\}
\end{equation}
where
\begin{equation}
\overline{Q}=\lim_{T \to \infty}\frac{1}{T}\sum_{t=0}^{T-1}\left\{\sum_{k\in \mathcal{K}}Z_k(t)+\sum_{k\in \mathcal{K}}\sum_{i\in M_k}[Q_i(t)+Q_{i,k}(t)]\right\}.
\end{equation}

\section*{Acknowledgment}

The work was supported in part by the National Key Research and Development Program of China (No. 2019YFE0111600), National Natural Science Foundation of China (No. 61971083, No. 51939001 and No. 62371085), LiaoNing Revitalization Talents Program (No. XLYC2002078) and Fundamental Research Funds for the Central Universities (No. 3132023514).

\ifCLASSOPTIONcaptionsoff
  \newpage
\fi

\end{document}